\documentclass[acmlarge]{acmart}

\AtBeginDocument{
  \providecommand\BibTeX{{
    \normalfont B\kern-0.5em{\scshape i\kern-0.25em b}\kern-0.8em\TeX}}}
    
\setcopyright{acmcopyright}
\copyrightyear{2023}
\acmYear{2023}
\acmDOI{XXXXXXX.XXXXXXX}

\acmJournal{CSUR}

\usepackage{amsmath,amssymb,amsfonts}
\usepackage{algorithmic}
\usepackage{graphicx}
\usepackage{textcomp}
\usepackage{xcolor}
\usepackage{tcolorbox}
\usepackage{framed}
\usepackage{array}
\usepackage{multirow}
\usepackage{mathrsfs}
\usepackage{color}
\usepackage{threeparttable}
\usepackage{subcaption}
\usepackage{hyperref}

\hypersetup{hidelinks,
	colorlinks=true,
	allcolors=black,
	pdfstartview=Fit,
	breaklinks=true}

\begin{document}

\title{Prompting Frameworks for Large Language Models: A Survey}

\author{Xiaoxia Liu}
\affiliation{
  \institution{Zhejiang University}
  \country{China}
}
\email{liuxiaoxia@zju.edu.cn}

\author{Jingyi Wang}
\affiliation{
  \institution{Zhejiang University}
  \country{China}
}
\email{wangjyee@zju.edu.cn}

\author{Jun Sun}
\affiliation{
  \institution{Singapore Management University}
  \country{Singapore}
}
\email{junsun@smu.edu.sg}

\author{Xiaohan Yuan}
\affiliation{
  \institution{Zhejiang University}
  \country{China}
}
\email{22332075@zju.edu.cn}

\author{Guoliang Dong}
\affiliation{
  \institution{Singapore Management University}
  \country{Singapore}
}
\email{gldong@smu.edu.sg}

\author{Peng Di}
\affiliation{
  \institution{Ant Group}
  \country{China}
}
\email{dipeng.dp@antgroup.com}

\author{Wenhai Wang}
\affiliation{
  \institution{Zhejiang University}
  \country{China}
}
\email{zdzzlab@zju.edu.cn}

\author{Dongxia Wang*}
\thanks{*Corresponding author}
\affiliation{
  \institution{Zhejiang University}
  \country{China}
}
\email{dxwang@zju.edu.cn}

\begin{abstract}
    Since the launch of ChatGPT, a powerful AI Chatbot developed by OpenAI, large language models (LLMs) have made significant advancements in both academia and industry, bringing about a fundamental engineering paradigm shift in many areas. While LLMs are powerful, it is also crucial to best use their power where ``prompt'' plays a core role.  
    However, the booming LLMs themselves, including excellent APIs like ChatGPT, have several inherent limitations: 1) temporal lag of training data, and 2) the lack of physical capabilities to perform external actions. 
    Recently, we have observed the trend of utilizing prompt-based tools to better utilize the power of LLMs for downstream tasks, but a lack of systematic literature and standardized terminology, partly due to the rapid evolution of this field. Therefore, in this work, we survey related prompting tools and promote the concept of the ``Prompting Framework" (PF), i.e. the framework for managing, simplifying, and facilitating interaction with large language models. 
    We define the lifecycle of the PF as a hierarchical structure, from bottom to top, namely: Data Level, Base Level, Execute Level, and Service Level.
    We also systematically depict the overall landscape of the emerging PF field and discuss potential future research and challenges. To continuously track the developments in this area, we maintain a repository at \href{https://github.com/lxx0628/Prompting-Framework-Survey}{https://github.com/lxx0628/Prompting-Framework-Survey}, which can be a useful resource sharing platform for both academic and industry in this field.
\end{abstract}



\maketitle

\section{introduction}
Since the release of ChatGPT \footnote{https://openai.com/blog/chatgpt/}, which attracted widespread social attention, research on large language models (LLMs) has been in full swing in both academia and industry, resulting in a number of amazing products such as PaLM \cite{chowdhery2022palm}, GPT-4 \cite{openai2023gpt4}, and LLaMA \cite{touvron2023llama, touvron2023llama2}. These LLMs have been shown to exhibit remarkable capabilities in approaching or even exceeding human-level performance in dialogue, text translation, and sentiment analysis \cite{bommasani2023holistic, chen2023robust, amin2023will, laskar2023systematic}, etc, potentially bringing in fundamental changes of many fields \cite{montenegro2023impact, zhou2023exploring, dwivedi2023so, liu2023using, lund2023chatgpt, cascella2023evaluating, xiao2023evaluating, dai2023can}. 

The development of language models to the current flourishing state has undergone a series of evolutionary processes: \emph{fully supervised learning $\to$ deep learning for NLP $\to$ ``Pre-train, Fine-tune" $\to$ ``Pre-train, Prompt, Predict" \cite{liu2023pre, zhao2023survey}}. 
Initially, language models (LMs) applied a fully supervised learning paradigm, where task-specific models were trained solely on the target task dataset, heavily relying on feature engineering \cite{lafferty2001conditional, och2004smorgasbord, rosenfeld2000two}. Subsequently, with the rise of deep learning, neural networks for NLP emerged, enabling the integration of feature learning and model training, i.e., a network architecture designed to automatically learn data features \cite{bengio2000neural, mikolov2010recurrent, collobert2011natural, bengio2013representation}. Later, as the demand for LMs increased and to accommodate the growing number of NLP tasks, the ``Pre-train, Fine-tune" paradigm was introduced. In this paradigm, a model with a fixed architecture undergoes pre-training to predict the probability of observed text data. Additional parameters are then introduced, and the model is fine-tuned using task-specific objective functions to adapt the pre-trained LM to various downstream tasks \cite{sarzynska2021detecting, yang2019xlnet, lewis2019bart, vaswani2017attention}. Then came the era of LLMs, where the trend shifted towards downstream tasks actively adapting to pre-trained models. The paradigm of ``Pre-train, Prompt, Predict" became mainstream and prompts successfully empowering the LLMs to effortlessly tackle a wide range of complex and diverse tasks. By providing a suitable set of prompts, a single language model trained entirely on context-based predictions can be employed to address various tasks \cite{brown2020language, raffel2020exploring}. Therefore, the quality and appropriateness of prompts are increasingly playing a crucial role in task resolution \cite{wei2022chain, zhou2022least, kojima2022large}. Both the academic and industrial communities have shown growing attention and interest in research related to prompts.

Numerous studies have demonstrated the necessity of employing appropriate methods to unleash the potential of LLMs \cite{wei2022chain, yao2023tree, wang2022self, zhou2022least}. In March 2023, OpenAI officially unveiled a significant innovation known as ChatGPT plugins, which enable ChatGPT to utilize external tools, reflecting a clear response to the growing demand for enhancing LLMs' interaction capabilities with the external world. When analogized to humans, LLMs can be regarded as the intelligent system's brain, responsible for perceiving instructions and generating and controlling a series of actions. Therefore, by combining their inherent knowledge and capabilities with external tools such as search engines, computational utilities, visual models, and more, LLMs can perform a wide array of real-world tasks, including real-time data retrieval, browser-based information retrieval, database access, precise mathematical calculations, complex language generation, and image analysis, thus showcasing their potential across diverse domains like education, healthcare, social media, finance, and natural sciences \cite{qin2023tool, nakano2021webgpt, mialon2023augmented, lu2023chameleon}. Consequently, \emph{the development of tools that facilitate the optimization and streamlining of the interaction process becomes crucial.} In this paper, we collectively refer to these forward-looking tools as a proposed novel concept: ``Prompting Framework" (PF).

In general, Prompting Framework is the upper layer which enables LLMs to interact with the external world.
A prompting framework \emph{manages, simplifies, and facilitates such interactions}, helping 
LLMs overcome fundamental challenges like data lag or ``brain in a vet''.
Moreover, prompting frameworks also serve as the basic infrastructure of recently emerging autonomous agents based on LLMs, such as AutoGPT \cite{Significant_Gravitas_Auto-GPT}, HuggingGPT \cite{Wolf_Transformers_State-of-the-Art_Natural_2020}, and MetaGPT \cite{hong2023metagpt}.


Since the release of the open-source project LangChain \cite{Chase_LangChain_2022} by Harrison Chase in October 2022, it has garnered attention from over 60,000 supporters on GitHub and stands as one of the most popular prompting frameworks to date. LangChain is a framework for building applications with LLMs through composability. Besides LangChain, our investigation encompasses various kinds of state-of-the-art prompting frameworks, including 1) Semantic Kernel \cite{Semantic-kernel}, LlamaIndex \cite{Liu_LlamaIndex_2022}, and OpenDAN \cite{OpenDAN.ai}, which can be arguably considered as the operating systems for LLMs, as well as 2) output restrictors for LLMs such as Guidance \cite{guidance}, TypeChat \cite{TypeChat}, NeMo-Guardrails \cite{NeMo-Guardrails}, and 3) language for interacting with LLMs, such as LMQL \cite{beurer2023prompting}, gpt-jargon \cite{gpt-jargon}, SudoLang \cite{SudoLang}. 
When referring to prompting frameworks, a notable challenge arises due to the rapid pace of development in the domain, making it difficult to track and stay informed about the multitude of methods dispersed across GitHub, preprint papers, Twitter, and top conferences/journals. Furthermore, the abundance of prompting framework approaches with varying focuses makes it challenging to systematically categorize and compare them, hindering the selection of the most suitable product for specific needs. Therefore, there is currently a lack of but an urgent need of systematic literature and standardized terminology for introducing and comparing these tools that are essential for better using LLMs' capabilities. 

In this survey, we introduce the concept of `Prompting Framework',
and provide a comprehensive and systematic survey of existing prompting frameworks. We present categorization, comparative analysis, and evaluation criteria for them, assess their applicability and limitations, and provide practical recommendations for their effective utilization for real-world LLM-enabled tasks. Additionally, we discuss some useful toolkits related to prompts that fall beyond the scope of Prompting Frameworks. We also present recommendations for future research. In a nutshell, we make the following main contributions:

\begin{itemize}
\item We introduce the concept of Prompting Frameworks that garnered attention in both academia and industry, and provided systematic and standardized definitions and terminology.
\item We categorize the existing Prompting Frameworks into 3 classes, conduct a comprehensive comparison of their strengths and limitations across various dimensions, and provide practical recommendations. Based on the research findings, we present the future directions of the Prompting Framework and extensively explore its potential development and challenges in more domains.
\item We conduct extensive research beyond the scope of prompting frameworks, including works and tools related to LLMs' prompts and task execution of prompting frameworks. We put them together in our GitHub repository to facilitate researchers' access and exploration for further studies.
\end{itemize}

The rest of the article is structured as follows. Section \ref{sec:background} presents background knowledge of the Prompting Framework, including the characteristics of LLMs and the necessity of the Propmpting Framework.
Section \ref{sec:over} describes the investigation, including the methodologies and results. Section \ref{sec:main} provides the systematic definitions and taxonomy of Prompting Frameworks. Section \ref{sec:com} presents the comparison and challenges across various dimensions of various Prompting Frameworks. 
Section \ref{sec:othertool} reviews prompt-based work outside the scope of the Prompting Framework but related to LLMs. 
Section \ref{sec:con} presents the future directions of the Prompting Framework and the potential developments and challenges in more domains.

\section{Background}
\label{sec:background}
In this section, we present the background of the Prompting Framework, including the reasons behind its emergence and the pertinent terminologies. We aim to address the following aspects: 1) elucidating the concept of LLMs by tracing their development history, and 2) explicating the current capability limitations of LLMs to underscore the necessity for the Prompting Framework.

\begin{figure*}[h]
  \centering
  \includegraphics[width=0.85\linewidth]{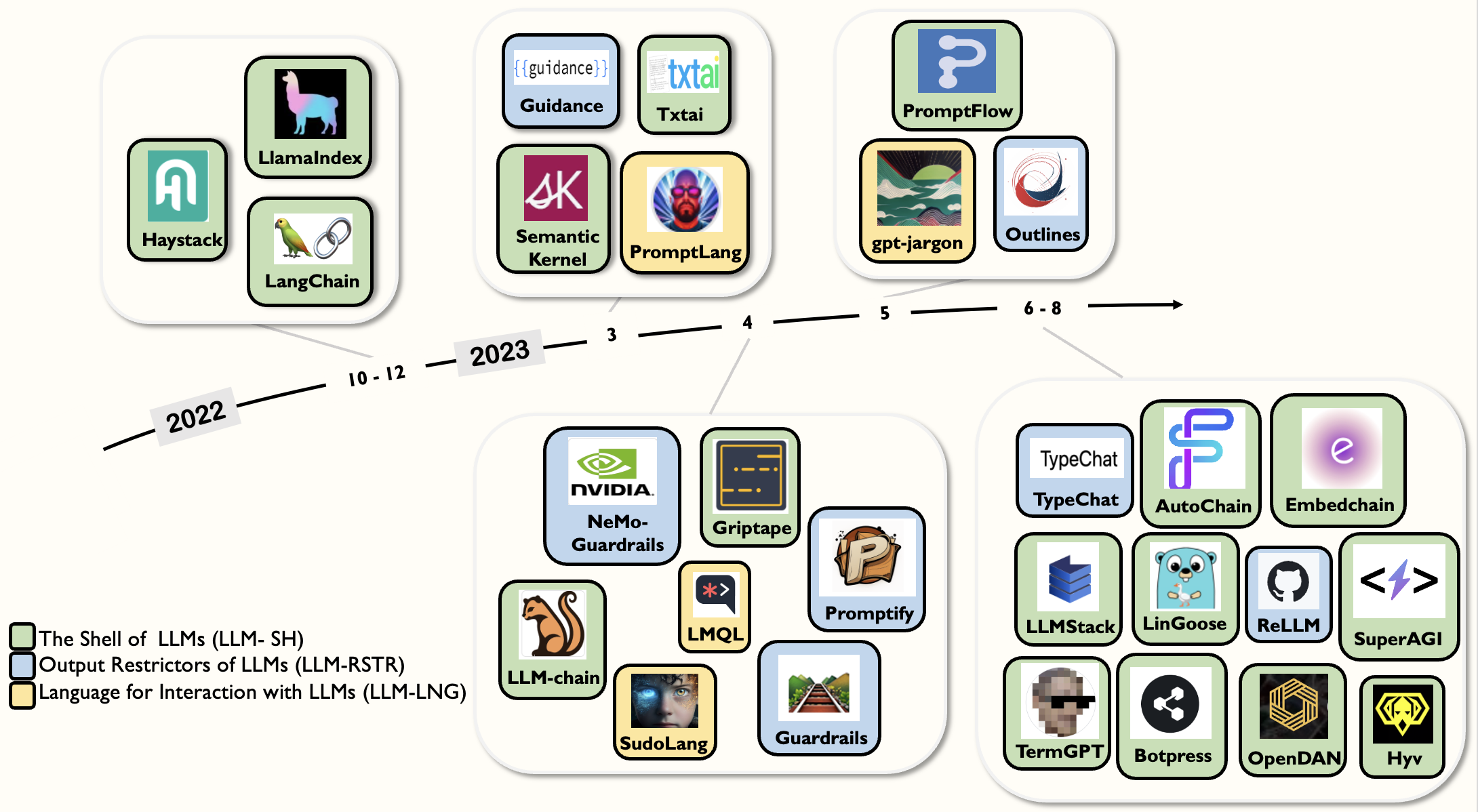}
  \caption{The timeline of representative prompting frameworks.}
  \label{sys-overview}
\end{figure*}

\subsection{Trends in Language Model: from LMs to $LLMs$}
Early language models were predictive models based on Markov assumptions using statistical learning methods, also known as Statistical language models (SLM) \cite{jelinek1998statistical, thede1999second, kotsiantis2007supervised, och2004smorgasbord}. However, due to the limitations imposed by the fully supervised learning approach, the curse of dimensionality was inevitably a challenge. With the rise of deep learning, researchers turned to neural networks to enhance LM's capabilities, leading to the emergence of Neural Language Models (NLMs) \cite{mikolov2013efficient, mikolov2013distributed}. NLMs aims to establish a universal neural network framework for various natural language processing (NLP) tasks. Subsequently, the introduction and popularity of the Transformer architecture and self-attention mechanism \cite{vaswani2017attention} gave rise to a series of task-agnostic pre-trained models, such as BERT and GPT, called Pre-trained language models (PLM), which promoted the emergence of the ``pre-train, fine-tune" paradigm \cite{liu2023pre}. PLMs have exhibited remarkable performance improvements across a wide range of NLP tasks \cite{devlin2018bert, lewis2019bart, liu2019roberta, radford2019language}.

To further explore the performance of LMs, researchers have continuously increased the scale of model parameters, the trend has shifted towards downstream tasks actively adapting to pre-trained models. The paradigm of ``Pre-train, Prompt, Predict" became mainstream \cite{liu2023pre}. Prompts are an important medium for interaction with Language Models, usually in text form. In this process, the augmented models not only exhibit better performance on various NLP tasks but also demonstrate remarkable ``emergent abilities" \cite{wei2022emergent}, which were previously unseen in smaller PLMs with similar architectures. For instance, ChatGPT can mimic human language style and logical reasoning and also demonstrates outstanding contextual comprehension, which was absent in previous models like GPT-2. Based on this new capability distinction, researchers refer to these emerging PLMs with hundreds of billions of parameters as Large Language Models (LLMs) \cite{wei2022chain, hu2021lora, kojima2022large}, such as ChatGPT, and GPT-4 \cite{openai2023gpt4}. With this advancement, language models have completed the leap from LMs to LLMs and inspiring new prospects for artificial general intelligence (AGI).

\subsection{LLMs still ``Brains in a Vat": Limitations and Mitigation}
Analogous to humans, LLMs can be perceived as the brains of artificial intelligence systems, responsible for perceiving instructional information and generating and controlling actions. 
Although there has been evidence of ``Emergent Abilities" \cite{wei2022emergent, webb2023emergent, yang2023harnessing, 10.1145/3605943} in LLMs, which refers to the abilities that emerge in large-scale models but have not been observed in smaller models and can be primarily categorized into four aspects: in-context learning\cite{dong2022survey, min2022rethinking, wei2023larger, wang2023images}, reasoning for complex content \cite{wei2022chain, lewkowycz2022solving, zhou2022least}, instruction following \cite{chung2022scaling, ouyang2022training, wang2022self, peng2023instruction}, and creative capacity \cite{cao2023comprehensive, zhang2023complete, zhang2023one, du2023enabling}. 

However, the capacity limitations of LLMs cannot be ignored. 
Firstly, LLMs suffer from temporal lag in their training data, such as ChatGPT's latest training data being limited to September 2021 (at the time of paper writing). LLMs are unable to access real-time information and trends, and they may struggle to accurately comprehend specific terminology or domain-specific knowledge, occasionally leading to incomplete or erroneous responses, even illusions \cite{chen2023robust, chang2023survey, li2023survey, wang2023robustness, cao2023assessing, deshpande2023toxicity, ferrara2023should}. Additionally, LLMs are bound by strict limitations on the number of tokens they can process during interactions, severely restricting the amount of contextual information they can take in and process \cite{yang2023exploring, zeng2023vcc, bulatov2023scaling, bertsch2023unlimiformer}. Secondly, LLMs are incapable of direct interaction with external expert models, such as utilizing search engines, querying databases, and invoking external tools, or APIs, which limits their usability \cite{qin2023tool, mialon2023augmented, zhao2023survey}. Furthermore, the majority of LLMs are offered as paid APIs, potentially imposing financial burdens on individuals, organizations, and projects with limited resources when dealing with large-scale or frequent requests \cite{chen2023frugalgpt, oppenlaender2023mapping}.

Based on the above challenges, it is thus desirable to overcome these barriers by
bridging the gap between LLMs and external applications. The adoption of Prompting Frameworks, such as Langchain and semantic kernel, becomes imperative \cite{montenegro2023impact, chen2023lm4hpc}. These frameworks not only enable LLMs to stay constantly exposed to emerging information but also enable the processing of long texts and documents, and facilitate seamless integration with external applications.

\section{survey overview}
\label{sec:over}
In this section, we provide a comprehensive description of our survey process.
The domains of LLMs and associated technologies are currently undergoing an unprecedented phase of rapid development. As a consequence, the landscape of relevant research and achievements is characterized by its dispersed nature. Many contributions have yet to be formally published in traditional academic journals or conferences. Instead, they are often found on platforms like arXiv or as open-source toolkits available on GitHub. Some noteworthy developments exist primarily within online communities on platforms such as Twitter, GitHub, and Discord, lacking formal documentation. Furthermore, there is a notable absence of comprehensive review literature in the field, resulting in a scarcity of established academic terminology and official definitions.

Our exploration of prompting frameworks begins with an in-depth examination of LangChain, recognized as one of the most influential frameworks in this domain. 
We start by delving into LangChain's official description, which emphasizes the concept of ``Building applications with Large Language Models (LLMs) through composability." This primary phase of our research seeks to establish a foundational understanding of the terminology and concepts central to these frameworks. We scrutinize and analyze terms such as ``frameworks," ``tools," ``Agent," ``Large Model," ``prompt," and ``toolkits." These keywords are thoughtfully selected to ensure an encompassing perspective, allowing us to include a wide range of relevant materials and resources.

In our pursuit of a comprehensive examination, we conduct multiple rounds of keyword searches across diverse platforms. This includes exhaustive searches on prominent repositories like GitHub and scholarly databases such as arXiv. Additionally, we extend our exploration to encompass reputable conferences and journals within the fields of artificial intelligence (AI) and natural language processing (NLP). These additional searches ensure that we are not only capturing the latest developments but also accessing academic and research-oriented materials of significance. Throughout this research process, our focus is to identify, collect, and analyze relevant materials. In total, we amass substantial works comprising 49 open-source projects available on GitHub and a significant number of academic papers. This methodical approach and rigorous examination of resources form the cornerstone of our research into prompting frameworks, facilitating a thorough and well-rounded exploration. 

Subsequently, our investigation delves into a meticulous and systematic assessment of the 49 works under scrutiny. This comprehensive evaluation begins with an exhaustive review of their technical documentation, wherein we scrutinize the minutiae of each work's conceptual underpinnings, functional implementations, and crucial code segments. We embark on an in-depth exploration, configuring and pragmatically employing these tools to conduct a scientific and methodical analysis, evaluating their performance, efficiency, and applicability. In detail, we conduct extensive testing and research, which involve running all the test cases provided in the technical documentation and manually creating numerous more detailed test cases that better reflect real-world requirements. Following the fundamental procedures of software testing, we begin with unit testing of each individual module within the framework. Subsequently, we proceed to performance testing of modules assembled according to requirements and standards in complex applications, thus accomplishing integration testing. Finally, we conduct comprehensive system testing to validate and evaluate the capabilities claimed in these tasks, while also organizing aspects related to user experience.

Finally, this multi-faceted examination enables us to identify the merits and limitations of each work, providing us with a nuanced understanding of their capabilities and relevance to the overarching objectives of our survey. Following this rigorous assessment, we judiciously select approximately 30 works that not only conform to the conceptual prerequisites of the prompting framework but also stand out in the field. These selected works are chosen to be included in our survey to ensure a comprehensive and representative illustration of the burgeoning and dynamically evolving landscape of the prompting framework, which significantly shapes interactions between individuals and LLMs.

\begin{figure*}[h]
  \centering
  \includegraphics[width=0.88\linewidth]{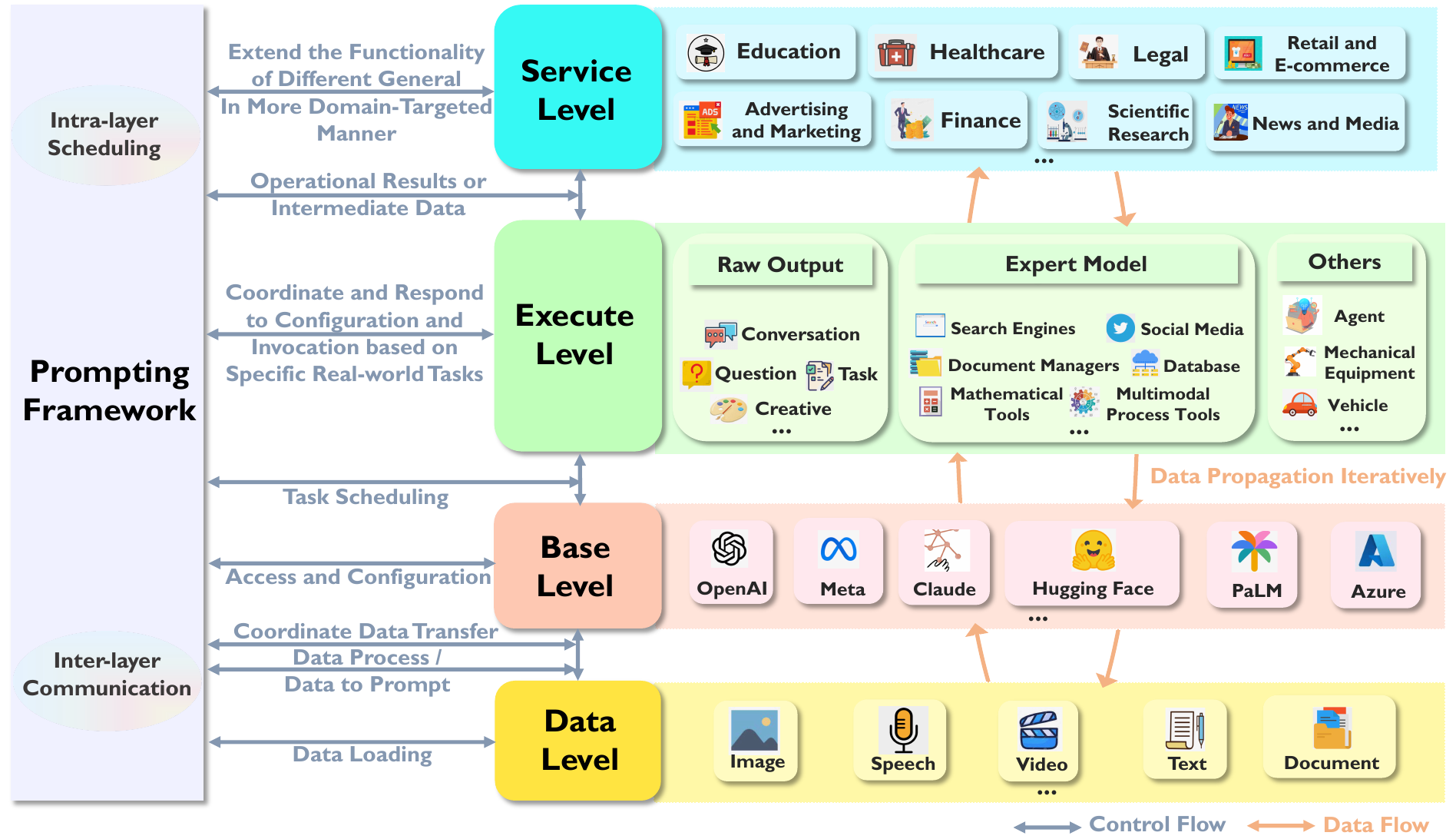}
  \caption{The workflow for facilitating interactions between LLMs and external entities using the Prompting Framework.}
  \label{fig-workflow}
\end{figure*}

\section{state-of-the-art prompting framework}
\label{sec:main}
In Sec.\ref{sec:background}, we analyze the current constraints and limitations of LLMs in practical applications. Despite these limitations, LLMs exhibit remarkable emergence abilities.
By combining LLMs with the Prompting Framework, the limitations of LLMs can be mitigated to some extent, enabling the realization of more astonishing capabilities. Therefore, in this section, we provide a systematic and comprehensive definition and description of the Prompting Framework, which is crucial to break down the application barriers of LLMs and empowers them as critical tools in real-world scenarios. We also classify the state-of-the-art Prompting Frameworks to provide a systematic review of various approaches. 

\subsection{Conceptualization}
Prompt typically serves as a crucial medium for interacting with LLMs, taking the form of textual content. A Framework is a general and extensible infrastructure that provides a structured approach and a set of guidelines. Consequently, we provide the following definition of the Prompting Framework: 
\begin{center}
\begin{tcolorbox}[colback=gray!13,
                  colframe=black,
                  width=\columnwidth,
                  arc=1mm, auto outer arc,
                  boxrule=0.35pt,
                 ]
\textit{\textbf{Prompting Framework (PF)}} is the framework for managing, simplifying, and facilitating interaction with large language models, which adheres to four essential properties: modularity, abstraction, extensibility, and standardization.
\end{tcolorbox}
\end{center}

\vspace{2mm}

Specifically, \emph{modularity} refers to breaking down the structure of the prompting framework into independent modules for easy code management and reusability; \emph{abstraction} refers to providing high-level, simplified interfaces to hide complex implementation details in the prompting framework's design; \emph{extensibility} inclines to allow users to customize and extend framework functionalities as needed, and \emph{standardization} refers to consistency in development to improve code maintainability and readability.

We categorize the workflow for facilitating interactions between LLMs and external entities using the Prompting Framework into four hierarchical layers, arranged from bottom to top as Data Level, Base Level, Execute Level, and Service Level. The Prompting Framework serves as the facilitator for inter-layer communication and intra-layer scheduling. It is important to note that the interactions between different levels are non-linear. In the course of a single task execution, data may iteratively propagate between various layers to accomplish intricate operations. The following is a detailed exposition of these four levels and the role played by the prompting framework within them:

\subsubsection{Data Level}
The Data Level is typically the foundational layer, serving as the most direct interface with the external environment. The Data Level primarily handles tasks such as data transmission and preprocessing, while being responsible for managing interactions with external data sources, such as databases or file systems. Within this process, the prompting framework plays a pivotal role in achieving a unified approach to dealing with various types of data, including text, images, videos, structured data, and documents. Simultaneously, the prompting framework has the capability to transform raw, unprocessed input into well-crafted prompts tailored to specific tasks or requirements, including question-answering, dialogue, and reasoning, facilitating more efficient and effective interactions with high-performance models at the Base Level.

\subsubsection{Base Level}
The Base Level operates as a computational hub situated between the data level and the execute level, serving as the analogical equivalent of the human brain or, in a computer analogy, the CPU. The Base Level primarily takes responsibility for the management of LLMs as the computational and control center, involving the reception and comprehension of instructions, execution of commands, and conducting various computations, which supports knowledge management and decision-making processes. Throughout this process, the prompting framework plays a critical role in coordinating data transfer and task scheduling. Furthermore, the prompting framework facilitates user-friendly access and flexible configuration of LLMs and can even autonomously select the most suitable LLMs for specific tasks.

\subsubsection{Execute Level}
The Execute Level constitutes a critical component of the business logic and is responsible for interacting with LLMs to accomplish specific real-world tasks. The Execute Level maintains communication with LLMs through the prompting framework collaboratively, and based on this, constructs tasks and takes appropriate actions based on the interactive information obtained from the prompting framework, and coordinates and responds to the configuration and invocation of models in alignment with LLMs to achieve the final completion of tasks. The Execute Level represents the terminal stage of task execution and primarily consists of three parts. The first part involves directly utilizing the Raw Output of LLMs from the Base Level to complete tasks without external assistance, representing the simplest business workflow. The second part entails selecting and invoking one or several external specialized models based on the interactive information from the prompting framework to handle aspects of task execution beyond the capabilities of LLMs, enabling the achievement of relatively complex tasks. The third part involves the coordination and integration of LLMs with higher-order models, such as interacting with various mechanical models (robotic arms, machines, vehicles, etc.) to realize LLM-based embodied intelligence or engaging with different types of agents to create LLM-based intelligent autonomous agents, further advancing the progress of Artificial General Intelligence (AGI).

\subsubsection{Service Level}
The Service Level resides at the top tier of the entire business workflow and is responsible for facilitating the management, scheduling, and integration of advanced tasks within specific domains, in coordination with the prompting framework. Service Level extends the functionality of different general in a more targeted manner by interacting with the prompting framework, particularly in critical domains such as education, healthcare, e-commerce, law, and finance.


\begin{table*}[t]
\caption{Representative Works of Prompting Frameworks.}
\label{tab:category}
\centering
\resizebox{.75\textwidth}{!}{%
\begin{tabular}{c|c|c}
\hline
Category & Subcategory & Representative Works \\ \hline
\multirow{2}{*}{\begin{tabular}[c]{@{}c@{}}The   Shell of LLMs \\ (LLM-SH)\end{tabular}} & Universal LLM-SH & \begin{tabular}[c]{@{}c@{}}Haystack \cite{Pietsch_Haystack_the_end-to-end_2019}\\ Semantic Kernel \cite{Semantic-kernel}\\\ LangChain \cite{Chase_LangChain_2022}\\ Griptape \cite{Griptape}\\ PromptFlow \cite{PromptFlow} \\ LLM-chain \cite{llm-chain}\\ LinGoose \cite{LinGoose}\\ LLMStack \cite{Chintala_LLMStack_A_platform}\\ OpenDAN \cite{OpenDAN.ai}\\ Hyv \cite{Hyv}\end{tabular} \\ \cline{2-3} 
 & Domain-Specific LLM-SH & \begin{tabular}[c]{@{}c@{}}LlamaIndex \cite{Liu_LlamaIndex_2022}\\ embedchain \cite{Singh_Embedchain_2023}\\ AgentVerse \cite{chen2023agentverse}\\ SuperAGI \cite{SuperAGI}\\ Txtai \cite{Mezzetti_txtai_the_all-in-one_2020}\\ AutoChain \cite{AutoChain}\\  TermGPT \cite{TermGPT}\\ Botpress \cite{Botpress}\end{tabular} \\ \hline
\multirow{2}{*}{\begin{tabular}[c]{@{}c@{}}Language   for Interaction with LLMs \\ (LLM-LNG)\end{tabular}} & Programming LLM-LNG & LQML \cite{beurer2023prompting}\\ \cline{2-3} 
 & Pseudocode LLM-LNG & \begin{tabular}[c]{@{}c@{}}PromptLang \cite{PromptLang}\\ SudoLang \cite{SudoLang}\\ gpt-jargon \cite{gpt-jargon}\end{tabular} \\ \hline
\multirow{2}{*}{\begin{tabular}[c]{@{}c@{}}Output   Restrictors of LLMs \\ (LLM-RSTR)\end{tabular}} & Content LLM-RSTR & \begin{tabular}[c]{@{}c@{}}NeMo-Guardrails \cite{NeMo-Guardrails}\\ Guardrails \cite{Guardrails}\end{tabular} \\ \cline{2-3} 
 & Structure LLM-RSTR & \begin{tabular}[c]{@{}c@{}}Guidance \cite{guidance}\\ Promptify \cite{Promptify2022}\\ ReLLM \cite{ReLLM}\\ TypeChat \cite{TypeChat}\end{tabular} \\ \hline
\end{tabular}
}
\end{table*}

\subsection{Taxonomy of Prompting Framework}
Taking into consideration the technical features, design objectives, and application scenarios, the current prompting framework can be broadly covered by three types: The Shell of LLMs (LLM-SH), Language for Interaction with LLMs (LLM-LNG), and Output Restrictors of LLMs (LLM-RSTR). In this section, we will elucidate the reasons for this classification and provide a detailed description of the characteristics and distinctions among these various types of prompting frameworks. The rationale behind designing the prompting framework is to facilitate the interaction between LLMs and the external world, and different types of prompting frameworks manifest this enhancement effect from different perspectives. LLM-SH functions are much like a shell or interface layer in computer systems, emphasizing interaction with LLMs by facilitating their engagement with highly capable third parties, thereby enabling stronger interaction between LLMs, users, and external models. LLM-LNG, on the other hand, is designed to create a language (programming or pseudo-language) for interaction with LLMs, focusing on providing users with a more concise and compact interaction channel. LLM-RSTR, meanwhile, achieves controlled generation by emphasizing interactions with LLMs that are of higher quality and better aligned with requirements. Furthermore, in the practical use of these tools, we have found that these three types of prompting frameworks are often compatible with each other. In other words, depending on the requirements, multiple different categories of prompting framework can be used in parallel within the same task-solving process.

\subsubsection{\textbf{The Shell of LLMs (LLM-SH)}} 
The shell of LLMs (LLM-SH) is a type of prompting framework aimed at enhancing the capabilities of LLMs by enabling them to access various external tools and knowledge sources. 
In traditional terms, a shell is often associated with an operating system, where the operating system's shell serves as a command-line interpreter that receives user input commands, interprets them, and passes them on to the operating system for execution, facilitating users in efficiently and comprehensively utilizing the functionality of the operating system. In other words, a shell can be viewed as a layer of encapsulation over the kernel, bridging the communication gap between commands and applications. Similarly, the design motivation behind LLM-SH, as a prompting framework, is to expand the action potential of LLMs. Specifically, with the assistance of LLM-SH, LLMs can not only accomplish conventional NLP tasks such as question answering, sentiment analysis, and information retrieval, but also higher-level functions across various domains including natural sciences, healthcare, education, finance, computer science, which encompass image processing and analysis, object detection, mathematical computations, database access, utilizing search engines, code comprehension and generation, social media posting, weather forecasting, and more. The LLM-SH simplifies complex interactions between LLMs and the external world, with a focus on improving their usability, universality, and scalability. LLM-SH also supports customization, allowing configuration and tailoring to specific needs and requirements for different application scenarios and specific domains.

In Tab. \ref{tab:category}, we provide classification and representative works according to categories. It is worth noting that we include many instances of ``Agent", but this paper primarily investigates the Agent framework rather than individual Agents. The distinction lies in the fact that LLMs-based Agents, exemplified by AutoGPT and BabyAGI \cite{BabyAGI}, emphasize the formulation of plans based on user-defined goals and the autonomous execution of these plans, whose main contributions lie in task acceptance, comprehension, automated decision-making, and execution processes. On the other hand, Agent frameworks, exemplified by SuperAGI \cite{SuperAGI} and AgentVerse \cite{chen2023agentverse} in the table, focus on customizing and building, managing, and running Autonomous Agents according to user-specific requirements. In simple terms, LLMs-based Agents are products primarily geared towards usage, whereas Agent frameworks serve as auxiliary tools to assist users in assembling and maintaining Agents, with an emphasis on construction.

There are two primary forms within LLM-SH. One is designed to support a wide range of applications and domains and is referred to as Universal LLM-SH. The other is more specialized and focused on a specific domain, such as Agent building and maintenance, data processing, or chat-bot building, and is known as Domain-Specific LLM-SH. 

\textbf{\textit{Universal LLM-SH}} 
is designed with the intention of accommodating a wide range of application scenarios and domains, offering extensive functionality and flexibility to meet the diverse needs of users, which typically possess higher generality and scalability.
Representative works include Haystack, Semantic Kernel, LangChain, Griptape \cite{Griptape}, PromptFlow \cite{PromptFlow}, LLM-chain \cite{llm-chain}, LinGoose \cite{LinGoose}, LLMStack \cite{Chintala_LLMStack_A_platform}, OpenDAN \cite{OpenDAN.ai}, Hyv \cite{Hyv}.

\textbf{\textit{Domain-Specific LLM-SH}} 
is specifically designed for particular domains or application scenarios. They are typically finely tuned and optimized for the requirements of that domain, often achieving better performance and higher efficiency in completing specific tasks.
Representative works include LlamaIndex \cite{Liu_LlamaIndex_2022} and Txtai \cite{Mezzetti_txtai_the_all-in-one_2020}, which are data frameworks designed to assist in building LLM-based applications. For building and managing LLM-based autonomous agents, we have AgentVerse \cite{chen2023agentverse} and SuperAGI \cite{SuperAGI}, along with AutoChain \cite{AutoChain}. In the domain of creating LLM-powered bots, there are embedchain \cite{Singh_Embedchain_2023} and botpress \cite{Botpress}. Additionally, for giving LLMs the capability to plan and execute terminal commands, there is TermGPT \cite{TermGPT}.

\subsubsection{\textbf{Language for Interaction with LLMs (LLM-LNG)}}
Language for Interaction with LLMs (LLM-LNG)  is an innovative type of prompting framework designed to facilitate more concise, direct, and compact interactions with LLMs by introducing a specialized language for programming.

Prompts serve as crucial intermediaries for interacting with LLMs and are typically presented in the form of natural language text. However, the capabilities of pure natural language text are limited and can increase complexity and cost when dealing with advanced tasks, sometimes even failing to yield accurate outputs. In contrast to purely natural language prompts, languages that integrate both natural language and programming logic are more structured, concise, and compact. They not only enhance reasoning performance but also provide better support for various prompting methods, such as chain-of-thought reasoning and decision trees. Additionally, they can offer improved support for control flow.

LLM-LNG comes in two primary forms: Programming LLM-LNG, which involves interactions with LLMs using programming languages, and Pseudocode LLM-LNG, which interacts with LLMs using a pseudo-code language. The main distinction between the two lies in their nature. Programming LLM-LNG belongs to the domain of programming languages and adheres to complex syntax rules and structures, which require compliance with specific syntax specifications to ensure program correctness and often require compilation or interpretation to transform it into executable code. On the other hand, Pseudocode LLM-LNG provides a simpler and more intuitive way of describing algorithms, combining natural language and structured coding with fewer formal constraints. As a result, Programming LLM-LNG, due to the interpretation and compilation process, tends to be more powerful in handling tasks and control flow. However, it also entails a steeper learning curve compared to Pseudocode LLM-LNG. Each approach has its own advantages.

\textbf{\textit{\textbf{Programming LLM-LNG}}} primarily involves designing a new programmable language for interacting with LLMs by simulating the syntax and architecture of existing programming languages. Given the significant overlap between user interactions with LLMs and the functionality of query languages, combining query language and language model prompts is a logical approach. This expansion transforms prompts from pure text-based prompts (natural language) to a combination of text prompts and scripts (natural language combined with programming language), enhancing the intuitiveness of interactions. In compound prompts that blend natural language with the programming language, constraints, and control flow are embedded into the instruction parsing and output parsing processes of LLMs through structured query languages. This functionality aims to streamline the reasoning process while reducing calls to resource-intensive underlying LLMs. Notable products in this category include LMQL \cite{beurer2023prompting}.

\textbf{\textit{Pseudocode LLM-LNG}} is a more open-ended form that flexibly combines natural language with structured coding, relying on the inherent capabilities of LLMs. Pseudocode is a wonderful method for outlining programs informally in natural language, without the constraints of specific syntax, which is like sketching out your ideas before diving into detailed coding. The design of Pseudocode LLM-LNG is driven by the need to unlock the full potential of LLMs, which possess strong capabilities but are hindered by inconvenient interactions or inaccurate prompts. Therefore, Pseudocode LLM-LNG offers a standardized structure and syntax that is easy to understand and interpret. It provides feasibility-verified template use cases or background explanations for communication with LLMs, combining natural language with simple coding conventions. Representative works in this category include PromptLang \cite{PromptLang}, SudoLang \cite{SudoLang}, and gpt-jargon \cite{gpt-jargon}.

\subsubsection{\textbf{Output Restrictors of LLMs (LLM-RSTR)}}
Output Restrictors of LLMs (LLM-RSTR) are a type of prompting framework designed to enable controlled generation by LLMs. The controlled generation problem with LLMs pertains to how to ensure that the generated text meets specific requirements, constraints, or demands, adapting to various application scenarios and professional domains. This involves control over multiple aspects such as semantic content, output structure, and semantic style. Currently, due to the uncontrolled nature of LLMs, the generated natural language text tends to be unstructured. Additionally, generated text may contain potential risks like bias, misinformation, or inappropriate content. LLMs also struggle with off-topic responses and maintaining consistency with predefined requirements. However, the application of LLM-RSTR can effectively alleviate these issues. LLM-RSTR primarily focuses on controlled generation from two perspectives: Content Control, referred to as Content LLM-RSTR, and Structure Control, referred to as Structure LLM-RSTR.

\textbf{\textit{Content LLM-RSTR}} focuses on achieving controlled generation by LLMs in three main aspects: privacy protection, security, and alignment with the topic and accuracy. As for privacy protection, Content LLM-RSTR ensures that user-provided personal or sensitive information is not leaked or misused. Security control aims to filter out unsafe, or dangerous content such as societal and cultural biases about gender, race, politics, and inappropriate or offensive content, and also prevents the generation of false or misleading information, thereby maintaining the accuracy and credibility of the generated content. The generated text should align with the user's or application's topic or requirements to ensure the generated content is useful. Additionally, text generation needs to maintain high accuracy, especially in specific domain applications such as medicine, law, or science. Representative works in this category include NeMo-Guardrails \cite{NeMo-Guardrails} and Guardrails \cite{Guardrails}.

\textbf{\textit{Structure LLM-RSTR}} plays a critical role in information processing, data management, and decision support, enabling both computers and humans to better understand and utilize the information, for tasks such as database management, search engine optimization, natural language processing (NLP), information extraction, data mining, and analysis. Unstructured original output from LLMs is challenging to use in business or other applications. Therefore, constraining and specifying the desired output text format is crucial. The most intuitive way to obtain structured text as output from LLMs is to write extensive and cumbersome tutorials and templates to instruct LLMs on what format the output should take. However, this is a time-consuming and complex process. The design of Structure LLM-RSTR aims to address this issue, allowing users to interact with LLMs in a simpler, more direct manner, and obtain structured outputs that are clear, easier to handle, and analyze. Representative works in this category include Guidance, Promptify \cite{Promptify2022}, ReLLM \cite{ReLLM}, and TypeChat \cite{TypeChat}.


\subsection{Crucial Component in the Construction of Prompting Framework}

After elucidating the concept of the prompting framework and conducting a comparative analysis, in this section,  we introduce one of the pivotal concerns within the prompting framework—specifically, the essential components required when constructing a prompting framework. Given the rapid development of both LLMs and the prompting framework itself, this field has not only given rise to numerous emerging technologies but has also reactivated many traditional techniques.

\subsubsection{\textbf{Vector Database}}
In the midst of the ongoing AIGC revolution, a particular challenge lies in the capability of large-scale storage and querying of unstructured data, such as images, videos, and text. Vector databases offer developers the means to handle unstructured data in the form of vector embeddings, which becomes especially crucial for the utilization and expansion of LLMs, for example, tools like OpenAI's Retrieval plugin rely on vector databases to assist users in retrieving relevant document excerpts from the data sources.
A vector database is a specialized database designed for storing and managing vector data. Vector data refers to data composed of multiple numerical values, often representing specific features or attributes. Vectorization is the process of transforming discrete variables, such as images and text, into continuous vector spaces. For example, different-sized or content images can be mapped into vectors within the same space, or various lengths of text can be mapped to a common vector space. In this space, adjacent vectors carry semantically similar meanings, and the vector space is commonly referred to as the embedding space, the generated vectors are known as embedding vectors or vector embeddings.
The primary characteristics of vector databases include efficient storage and querying of large-scale vector data. Typically, they employ queries based on vector similarity, retrieving data based on the similarity between vectors. This querying approach finds applications in various scenarios like image search, music recommendation, and text classification.

Vector databases depend on three key elements: vectorization (encoding), data structure, and distance calculation. The quality of vectorization determines the upper limit of vector database performance, yet current vectorization processes lack universality due to their strong dependence on data types. Properly constructing data structures to manage vectors ensures computational and retrieval efficiency, which determines the lower limit of vector database performance. Reasonable distance calculation between vectors can minimize resource consumption.

In recent years, there has been a proliferation of specialized database products. For instance, Milvus\cite{2021milvus} is considered the world's first true vector database product, with over 1000 enterprise users worldwide, making it one of the most popular open-source vector databases globally. Pinecone\cite{Pinecone}, designed for machine learning applications, offers speed, scalability, and support for various machine learning algorithms. Pinecone is also a partner of OpenAI, and users can generate language embeddings using OpenAI's Embedding API. Weaviate\cite{Dilocker_Weaviate}, a vector database, can store as many as billions of vectors. Additionally, Weaviate has introduced a Plug-in for ChatGPT, which has received recognition from OpenAI. The main distinction between Weaviate and Pinecone lies in how they manage services. Pinecone handles data storage and resource management fully for users, often in conjunction with AWS or GCP hosting. In contrast, Weaviate allows users to self-host their data while providing supportive operations and services. For users who value retaining control and not relinquishing their data entirely, Weaviate offers greater flexibility but may come with a relatively higher time cost.

\subsubsection{\textbf{Cache for LLMs}}
Fundamentally, every form of computation necessitates storage. Computation and storage represent the two fundamental abstractions, yet they are mutually convertible: storage can be exchanged for computation, and vice versa. Achieving an optimal trade-off is crucial in enhancing the input-output ratio. Whether dealing with large-scale or small-scale models, they fundamentally encode global knowledge and operational rules, serving as a compression of all human data. However, embedding all data into LLMs is challenging. For instance, some assert that ChatGPT serves as a highly efficient compression encoding, albeit not achieving lossless compression, in which the process inevitably introduces entropy reduction and information loss. Encoding all information into neural networks results in an excessively bulky model with an enormous parameter scale, leading to sluggish performance. Therefore, complete integration is unfeasible, implying the potential necessity for external storage. Similar situations exist in computer architecture, where the CPU incorporates on-chip SRAM, typically constrained in size due to the significantly higher cost of on-chip storage (100 times more expensive than DRAM and 10,000 times more expensive than disk storage). Neural networks function as on-chip storage for large models, with larger-scale models possessing more on-chip storage. However, utilizing neural networks for data storage proves costly, causing a rapid escalation in network scale. Hence, large models require a more efficient data storage method beyond neural networks, known as memory of LLMs.

For instance, GPTCache\cite{GPTCache} is a tool specifically designed to build semantic caches for storing LLMs' responses. It employs a modular design, including six main modules: LLM adapter, embedding generator, cache storage, vector store, cache manager, and similarity evaluator. The system offers multiple implementation options for each module, allowing users to customize their semantic cache to meet specific needs. Zep is a long-term memory store designed for building conversational LLM applications, which supports storing, summarizing, embedding, indexing, and enriching the history of LLM applications/chatbots. Zep\cite{Zep} enables long-term memory persistence, automatic summarization based on configurable message windows, vector search, and automatic memory token counting.

\subsection{Typical Applications of Prompting Framework}
In this section, we elucidate some typical applications of the prompting framework throughout the entire lifecycle of LLMs. These applications, situated at the closest proximity to the user at the application layer, are poised to offer boundless insights and inspiration for future developers and users.

\subsubsection{\textbf{Integrated Application Platform}}
The initial goal of constructing the prompting framework was to reduce the interaction barriers with LLMs and facilitate the development of LLM applications. Therefore, after addressing prototype issues, the subsequent challenge is to assist these applications in transitioning to practical development while ensuring implementation in a reliable and maintainable manner. Debugging, testing, evaluating, and monitoring the intricate data and control flow within LLM systems are crucial steps in this process, which ensures the robust deployment of LLMs in real-world production scenarios is of paramount importance. Consequently, considering both immediate requirements and future strategic objectives, integrated application platforms emerge as a typical application of the prompting framework.

For example, The newly developed LangSmith \cite{Chase_LangChain_2022} by LangChain developers introduces innovative features centered around five core pillars: debugging, testing, evaluation, monitoring, and usage metrics. LangSmith facilitates the execution of these operations through a simple and intuitive user interface, significantly lowering the barriers for developers without a software background. From a numerical perspective, many features of LLMs lack intuitiveness, making visual representation essential. We observe that a thoughtfully designed user interface can expedite user prototyping and work, as handling everything through code alone can be cumbersome. Furthermore, visualizing the processes and intricate command chains of LLM systems proves valuable in understanding the reasons behind specific outputs. As users construct more complex workflows, comprehending how queries traverse different processes becomes challenging. Therefore, a user-friendly interface to visualize these processes and record historical data represents a forward-looking innovative application.

\subsubsection{\textbf{LLM-based Agent}}
For a long time, autonomous agents have been a significant research topic. However, before the advent of LLMs and related technologies, limitations in training data, training methods, and interaction with the environment severely constrained the capabilities of agents. Consequently, agents struggled to make decisions similar to humans and achieve remarkable performance. However, with the current prevalence of LLMs and their outstanding capabilities, LLM-based autonomous agents have demonstrated immense potential in task processing and autonomous decision-making.

In a broad sense, an agent refers to any system capable of thinking, interacting with the environment, operating independently, and collaborating with other entities. In theory, given any objective, an agent should be able to achieve it automatically. LLM-based Agents belong to AI systems that autonomously generate sub-agents, which can execute tasks independently based on user requirements without the user's direct intervention, following the basic three sub-steps used by people to solve various problems: perception, decision-making, and action. LLM-based Agents can handle tasks ranging from daily event analysis, marketing plan generation, and code programming, to mathematical computations, among others. If ChatGPT follows user instructions, doing what the user tells it to do, then an LLM-based Agent acts on what it deems should be done. In other words, LLM-based Agents demonstrate a potential form of integration between LLMs and prompting frameworks. However, it's important to note that this is still an experimental concept and not a fully realized commercial product. Currently, LLM-based autonomous agents typically follow a unified architecture, consisting of four main modules: a configuration module representing agent attributes, a memory module for storing historical information, a planning module for formulating future action strategies, and an action module for executing plan decisions.

AutoGPT \cite{Significant_Gravitas_Auto-GPT}, released on GitHub by Significant Gravitas, is a well-known autonomous agent capable of executing actions based on LLMs' autonomous decision results and external resources. AutoGPT uses a cyclic evaluation strategy to assess the degree of goal achievement in real-time, determining whether a task is complete. AutoGPT is mainly composed of three parts: task distribution, autonomous execution, and result output. The autonomous execution module is the core of AutoGPT. Currently, AutoGPT can perform basic tasks such as internet searches and information collection, long-term and short-term memory management, access to common websites and platforms, and extension through plugins. HuggingGPT \cite{Wolf_Transformers_State-of-the-Art_Natural_2020}, developed by Zhejiang University and Microsoft Research Asia, is a collaborative system that connects LLMs with the ML community (HuggingFace). It can handle inputs from various modalities and address a wide range of complex AI tasks. In essence, HuggingGPT takes introductions to all models on the HuggingFace community as input and runs them through models. Then, based on the user's input question, it parses matches and decides which model to use for solving the task. Similarly, HuggingGPT's workflow comprises four stages: task planning, model selection, task execution, and response generation, which aligns closely with that of AutoGPT. In addition, AgentGPT \cite{AgentGPT} is a web-based solution that allows for the configuration and deployment of autonomous AI agents, facilitating interactive experiences with web users. CAMEL \cite{CAMEL}, short for ``Communicative Agents for 'Mind' Exploration of Large Scale Language Models," implements a novel role-playing agent. GPTRPG \cite{GPTRPG} combines game design with large language models, enabling the deployment of multiple agents to autonomously participate in online games by embedding AI agents into the roles within the game environment using the OpenAI API.

\section{Related Prompting Tools}
\label{sec:othertool}
In this section, we provide an extensive overview of prominent prompting tools that contribute to generating higher-quality prompts or achieving advanced functionality through prompts. Since these tools do note possess all four fundamental characteristics of prompting frameworks, namely modularity, abstraction, scalability, and standardization, they are not classified as prompting frameworks. Nevertheless, the problems they address and the functionalities they enable are also significant for future interactions with LLMs. Additionally, we introduce some auxiliary tools that play a vital role in task completion of prompting frameworks. 
Links to these tools are also organized in our GitHub repository.

Prompt is a powerful tool that enhances the flexibility and controllability of large language models (LLMs), making them applicable across various domains. Through clever design and utilization of prompts, users can guide the model to generate desired text, resulting in improved performance and effectiveness across various tasks. It can be stated that a well-crafted prompt can significantly boost the productivity of LLMs. The significance of predefined prompt templates, example libraries, or prompt optimization tools that template user-inputted prompts for large language models lies in their ability to not only lower technical barriers, enabling non-technical individuals to easily interact with the model, but also enhance interaction efficiency. In other words, users can select suitable prompts from existing templates without the need to write their own from scratch. Furthermore, since templates are designed and tested, they tend to be more precise and reliable, reducing errors caused by unclear or vague user instructions.

\textbf{Prompt's Template Library.} Awesome ChatGPT Prompts \cite{awesome-chatgpt-prompts} is an open-source website and application created by JavaScript developer Fatih Kadir Akın, which contains over 160 prompt templates for ChatGPT, allowing it to mimic a Linux terminal, JavaScript console, Excel page, and more. These prompts have been collected from excellent real-world use cases. LangGPT \cite{LangGPT} is designed to write high-quality prompts in a structured, templated manner. It not only provides templates but also supports variable configuration and references based on templates. PromptSource \cite{Bach_PromptSource_An_Integrated_2022} is a toolkit for creating, sharing, and using natural language prompts. PromptSource allows for the use of thousands of existing and newly created prompts, which are stored in separate structured files and written in a simple template language called Jinja.

\textbf{Optimizer for Prompts.} OpenPrompt \cite{Ding_OpenPrompt_An_Open-source_2021} provides a standardized, flexible, and extensible framework for deploying prompt-based learning pipelines. It supports existing prompt learning methods and allows for the design of custom prompt learning tasks. HumanPrompt \cite{humanprompt} is a framework that makes it easier for humans to design, manage, share, and use prompts and prompting methods. InstructZero \cite{chen2023instructzero} aims to optimize poorly phrased prompts provided by users to LLMs, transforming them into well-structured and compliant prompts. The optimization process primarily aligns humans with LLMs, rather than fine-tuning LLMs to align with humans, as in instruction fine-tuning.



\textbf{Evaluation of LLMs.} Evaluating LLMs has always been a crucial topic to ensure their reliability, safety, usability, and compliance while helping identify potential issues and improvement areas \cite{bang2023multitask, jain2023bring, lin2023llm, qin2023chatgpt}. Evals \cite{Evals} is a framework for evaluating LLMs and LLM systems and serves as an open-source registry of benchmarks. Evals simplifies the process of constructing evaluations with minimal code while being straightforward to use. PromptBench \cite{zhu2023promptbench} is a framework for robustness evaluations of large language models under adversarial prompts, which facilitates examining and analyzing interactions between large language models and various prompts, providing a convenient infrastructure for simulating black-box adversarial prompt attacks and evaluating performance. PromptInject \cite{ignore_previous_prompt} is a framework that modularly assembles prompts to provide quantitative analysis of LLMs' robustness against adversarial prompt attacks.

\begin{figure*}[h]
  \centering
  \includegraphics[width=.95\linewidth]{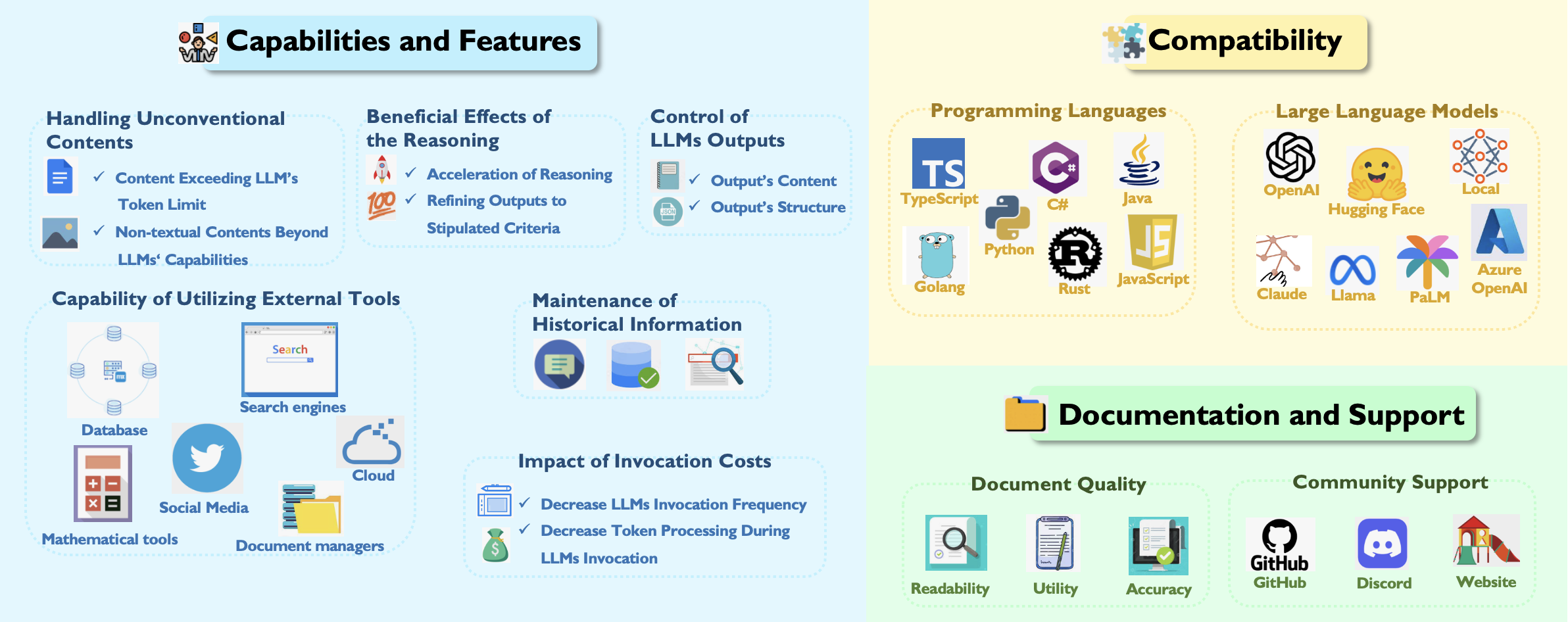}
  \caption{The dimensions and metrics for comparative analysis.}
  \label{fig-dimension}
\end{figure*}

\section{comparisons and challenges}
\label{sec:com}
\begin{figure*}[h]
  \centering
  \includegraphics[width=\linewidth]{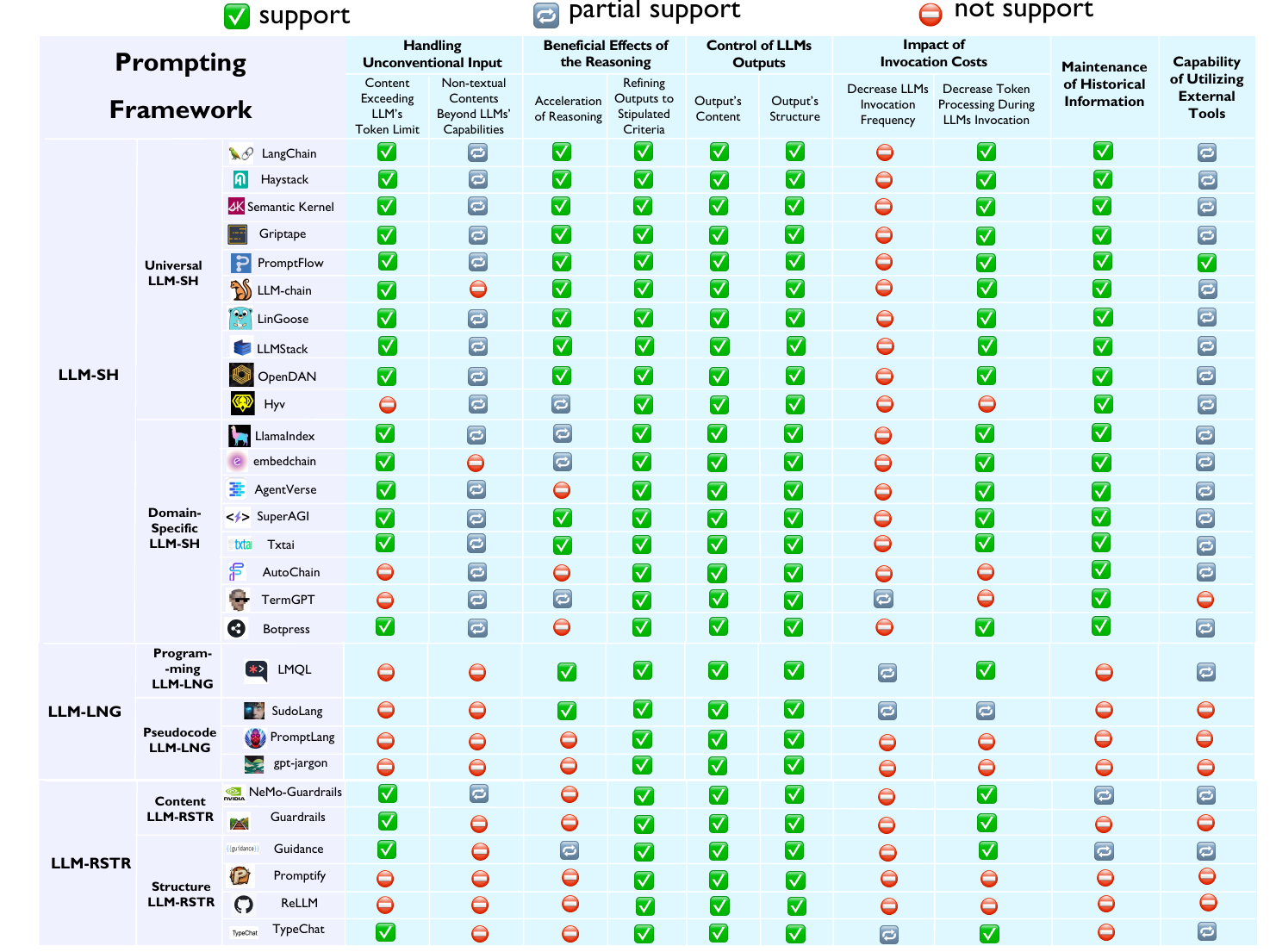}
  \caption{The capability matrix of representative prompting frameworks.}
  \label{ab-matrix}
\end{figure*}

In this section, we provide a comparison of existing prompting frameworks from various dimensions and analyze the challenges that prompting frameworks encounter in terms of development, practical implementation, and further advancements. The dimensions and metrics for conducting the comparative analysis are shown in Fig. \ref{fig-dimension}, and the detailed capability matrix based on the dimensions and metrics above of the mainstream prompting framework is illustrated in Fig. \ref{ab-matrix}.

\subsection{Comparative Analysis of Prompting Frameworks}
We have examined the three macro dimensions of compatibility, capabilities and features, as well as documentation and support. Within these dimensions, we have focused on more detailed key issues, providing a comprehensive analysis and comparison of existing prompting frameworks, which offer systematic experiences and guidelines for developers and users in their practical adoption of prompting frameworks and for further advancements. 


\subsubsection{\textbf{Compatibility}}
Compatibility refers to the adaptability and interoperability of systems, software, hardware, or other components in various environments or conditions.  We primarily investigate the compatibility of the prompting framework with programming languages and its compatibility with LLMs.


\vspace{2mm}
\textbf{Compatibility of Programming Languages. }
Due to different development team preferences and project requirements, prompting frameworks are typically designed to consider the use of multiple programming languages by developers. As a result, various prompting frameworks often provide interfaces supporting one or multiple mainstream programming languages to allow developers to flexibly choose and interact with languages in different projects and environments, thereby enhancing development efficiency and adaptability. The currently available prompting framework provides comprehensive coverage of programming languages, including Python, Java, JavaScript, C\#, C++, Go (Golang), Rust, and TypeScript.

\textit{Comparative Analysis.}
Currently, the prompting framework is still in its early stages of rapid development. Language support tends to begin with an implementation in a specific language and then expand to support a wider range of programming languages, accommodating more use cases and domains. Overall, LLM-SH is designed to be more adaptable to a broader range of scenarios and to support a richer set of tools. Consequently, LLM-SH exhibits stronger compatibility with programming languages compared to LLM-RSTR and LLM-LNG. For instance, Txtai, within LLM-SH, offers impressive support for five mainstream programming languages: Python, Java, Rust, Golang, and JavaScript. Meanwhile, LangChain and Semantic Kernel can each support three programming languages: Python, TypeScript, and JavaScript, as well as C\#, Python, and Java, respectively.
Most prompting frameworks tend to focus on supporting one mainstream programming language, with Python and TypeScript being the primary choices. For instance, in LLM-SH, tools like Haystack, Griptape, PromptFlow, LLMStack, OpenDAN, AutoChain, TermGPT, and in LLM-LNG, LMQL, as well as in LLM-RSTR, NeMo-Guardrails, Guardrails, Guidance, Promptify, and ReLLM primarily support Python. On the other hand, Hyv and botpress belonging to LLM-SH, and TypeChat belonging to LLM-RSTR are specifically designed for TypeScript. LlamaIndex and embedchain in LLM-SH support both Python and TypeScript.
Furthermore, LLM-chain is exclusively focused on Rust, while LinGoose is tailored for the Golang. As for Pseudocode LLM-LNG, its intent is to create a new language to simplify interaction with LLMs, so SudoLang, PromptLang, and gpt-jargon support only the syntax of the language designed within the framework and natural language.
      

\begin{center}
\begin{tcolorbox}[colback=gray!13,
                  colframe=black,
                  width=\columnwidth,
                  arc=1mm, auto outer arc,
                  boxrule=0.35pt,
                 ]
\textit{In summary, in terms of compatibility with programming languages, LLM-SH surpasses LLM-RSTR, which in turn surpasses LLM-LNG.}
\end{tcolorbox}
\end{center}



\vspace{2mm}
\textbf{Compatibility of LLMs.}
The original intention of the Prompting Framework is to establish a medium for interaction between external and LLMs. Therefore, one of the key requirements for the Prompting Framework is its compatibility with LLMs, which means that the Prompting Framework should seamlessly integrate with various types of LLMs and handle their inputs and outputs correctly. In other words, Prompting Frameworks with better compatibility generally offer a unified interface, which enables users to use the different Prompting Frameworks with corresponding instructions to interact with LLMs,  without requiring additional handling of specific details. The current prompting framework supports various LLMs, primarily including the previously mentioned OpenAI API, Azure OpenAI API, HuggingFace API, as well as other APIs such as Llama API, Anthropic, ERNIE-Bot, Google PaLM, and so on.

\textit{Comparative Analysis.}
Currently, support for LLMs within prompting frameworks is in a phase of rapid evolution due to the continuous emergence of new LLMs. Developers need to swiftly integrate these models into prompting frameworks once they are familiar with their functionalities and usage. It's worth noting that when we refer to LLMs compatibility, we mean the ability to seamlessly integrate models into prompting frameworks without the need for complex additional operations or coding, which implies direct utilization of models in prompting frameworks through modifications to certain configurations, using the LLMs in their native supported form, rather than relying on additional plugins or frameworks. Models that are natively supported by prompting frameworks tend to integrate better with the framework, fully leveraging the model's capabilities and minimizing unexpected bugs during programming. At present, all prompting frameworks offer good support for prominent APIs released by OpenAI, namely text-davinci-003, gpt-3.5-turbo, chatgpt, and gpt-4. In other words, prompting frameworks were originally introduced to facilitate user interactions with these mainstream APIs. Pseudocode LLM-LNG is a special case where this mutation-free pseudocode language can work well with any interactive interface-based LLMs. Consequently, SudoLang, PromptLang, and gpt-jargon can be compatible with almost all interactive LLMs.
Only a few prompting frameworks currently offer support for Hugging Face APIs. For instance, within LLM-SH, LangChain, Semantic Kernel, Haystack, Griptape, PromptFlow, LinGoose, LlamaIndex, embedchain, Txtai, AutoChain, and within LLM-RSTR, TypeChat and Promptify, provide such support. The mentioned prompting frameworks can also be compatible with Google's Azure OpenAI API. Furthermore, for other popular APIs like Claude from Anthropic, Llama API, Google PaLM, etc., only LLM-SH's LangChain, Semantic Kernel, Haystack, Griptape, Txtai, LlmaIndex, and embedchain can offer good support.


\begin{center}
\begin{tcolorbox}[colback=gray!13,
                  colframe=black,
                  width=\columnwidth,
                  arc=1mm, auto outer arc,
                  boxrule=0.35pt,
                 ]
\textit{In summary, all prompting frameworks offer good support for OpenAI's APIs. In general, LLM-SH exhibits the strongest compatibility with LLMs, while Pseudocode LLM-LNG demonstrates exceptional compatibility when dealing with interactive interface-based LLMs without requiring compilation. }
\end{tcolorbox}
\end{center}


\subsubsection{\textbf{Capabilities and Features.}}
Capabilities and features are a crucial dimension when comparing different Prompting Frameworks, as they directly determine the framework's ability and flexibility in addressing problems and meeting user requirements. We elaborate on various crucial stages of the interaction between LLMs and the Prompting Framework, including data preprocessing, reasoning process, output control, cost considerations, tool learning, and information maintenance. The comparison dimensions of capabilities and features we have enumerated can be employed not only for analyzing the strengths and limitations of diverse prompting frameworks but also as evaluation metrics for forthcoming tasks in assessing the prompting frameworks.

\begin{table}[]
\centering
\caption{The maximum tokens supported and pricing of mainstream LLMs.}
\label{tab:cost}
\resizebox{0.65\textwidth}{!}{%
\begin{tabular}{|c|c|c|cc|}
\hline
\multirow{2}{*}{Model} & \multirow{2}{*}{Developer} & \multirow{2}{*}{Max Token} & \multicolumn{2}{c|}{Cost / 1K tokens} \\ \cline{4-5} 
 &  &  & \multicolumn{1}{c|}{Input} & Output \\ \hline
\multirow{2}{*}{gpt-4} & OpenAI & \multirow{2}{*}{8,192} & \multicolumn{1}{c|}{\multirow{2}{*}{\$0.03}} & \multirow{2}{*}{\$0.06} \\ \cline{2-2}
 & Azure   OpenAI &  & \multicolumn{1}{c|}{} &  \\ \hline
\multirow{2}{*}{gpt-4-32k} & OpenAI & \multirow{2}{*}{32,768} & \multicolumn{1}{c|}{\multirow{2}{*}{\$0.06}} & \multirow{2}{*}{\$0.12} \\ \cline{2-2}
 & Azure   OpenAI &  & \multicolumn{1}{c|}{} &  \\ \hline
\multirow{2}{*}{gpt-3.5-turbo} & OpenAI & \multirow{2}{*}{4,097} & \multicolumn{1}{c|}{\multirow{2}{*}{\$0.0015}} & \multirow{2}{*}{\$0.002} \\ \cline{2-2}
 & Azure   OpenAI &  & \multicolumn{1}{c|}{} &  \\ \hline
\multirow{2}{*}{gpt-3.5-turbo-16k} & OpenAI & \multirow{2}{*}{16,385} & \multicolumn{1}{c|}{\multirow{2}{*}{\$0.003}} & \multirow{2}{*}{\$0.004} \\ \cline{2-2}
 & Azure   OpenAI &  & \multicolumn{1}{c|}{} &  \\ \hline
\multirow{2}{*}{text-davinci-003} & OpenAI & \multirow{2}{*}{4,097} & \multicolumn{1}{c|}{\multirow{2}{*}{\$0.012}} & \multirow{2}{*}{\$0.016} \\ \cline{2-2}
 & Azure   OpenAI &  & \multicolumn{1}{c|}{} &  \\ \hline
\multirow{2}{*}{text-davinci-002} & OpenAI & \multirow{2}{*}{4,097} & \multicolumn{1}{c|}{\multirow{2}{*}{\$0.012}} & \multirow{2}{*}{\$0.012} \\ \cline{2-2}
 & Azure   OpenAI &  & \multicolumn{1}{c|}{} &  \\ \hline
text-embedding-ada-002 & OpenAI & 8,191 & \multicolumn{1}{c|}{\$0.0001} & \$0.0001 \\ \hline
Claude   Instant & Anthropic & 100,000 & \multicolumn{1}{c|}{\$0.00163} & \$0.00551 \\ \hline
Claude 2 & Anthropic & 100,000 & \multicolumn{1}{c|}{\$0.01102} & \$0.03268 \\ \hline
Llama   2 & MetaAI & 4,096 & \multicolumn{1}{c|}{free} & free \\ \hline
Cohere & Cohere & 4,096 & \multicolumn{1}{c|}{\$0.015} & \$0.015 \\ \hline
PaLM 2 & Google & 8,000 & \multicolumn{1}{c|}{\$0.0005} & \$0.0005 \\ \hline
\end{tabular}%
}
\end{table}

\vspace{2mm}
\textbf{Capability of Handling Unconventional Contents.}
The ability of prompting frameworks to assist LLMs in handling unconventional content primarily manifests in two aspects. Firstly, the prompting framework helps LLMs deal with content that exceeds the token limit, and secondly, the prompting framework empowers LLMs to process non-textual content formats that go beyond their inherent capabilities.
Usually, LLMs have a token limit determined by the model's architecture and memory constraints. For example, GPT-3.5-turbo and text-davinci-003 have a maximum input length of 4,097 tokens, while gpt-3.5-turbo-16k allows for 16,385 tokens, and GPT-4-32k allows for 32,768 tokens, which results in LLMs being unable to capture the global context of a text when dealing with extremely long documents without external support because only a portion of the text can be included. The maximum tokens supported and pricing of mainstream LLMs are shown in Tab. \ref{tab:cost}. Moreover, processing such long text can lead to performance issues, especially in resource-constrained environments. LLMs need to maintain a lot of information within a single input, which may require more computational resources and time. Therefore, prompting frameworks have emerged to assist LLMs in handling complex tasks involving extremely long texts, such as machine translation, legal document analysis, sentiment analysis of lengthy novels or articles, long-context dialogue systems, knowledge graph construction, etc.
LLMs are primarily designed for processing pure textual data, making them less suitable for unconventional formats like images, audio, or videos, whose main inputs and outputs are text data. Prompting frameworks effectively mitigate this limitation, enabling LLMs to handle a variety of data types, which broadens the application of LLMs to a wide range of multimedia tasks.

\textit{Comparative Analysis.}
In terms of the capability to handle content exceeding token limits and non-textual content, only LLM-SH can achieve these functionalities without the need for additional plugins or program calls. LLM-LNG and LLM-RSTR achieve these capabilities by making calls to LLM-SH within their framework design. However, it's important to note that such compatibility can potentially lead to unexpected bugs.

For handling content that exceeds token limits, LLM-SH empowers LLMs through methods like splitting, filtering, concatenation, or summarization. For instance, a classic approach is the ``Retrieval" module in LangChain, where ``Document transformers" offer pre-packaged functional functions for document splitting, composition, and filtering. In the ``Chains" module, there is a package called ``chains.summarize" that provides various methods for summarizing documents (PDFs, Notion pages, customer questions, etc.), including Map-Reduce, Stuff, and Refine approaches to organizing documents. By configuring parameters to design a "chain" structure differently and introducing vector databases and text embedding models, LLMs can effectively manage extremely long documents. Similarly, modules like "Summarizer" in Haystack and "Summary Engines" in GripTape can summarize long texts, and components like "PreProcessor" in Haystack and "Chunkers" in GripTape can perform text splitting and filtering for long texts.

When it comes to handling non-textual content, LLM-SH can process text extracted from non-text formats like YouTube videos or HTML files or generate multi-modal content such as images, videos, and audio, based on pure text content. However, it doesn't provide full-fledged processing of multimedia files throughout the input-output process. For instance, the "Tool Memory" module in GripTape can generate images, videos, PDFs, and other non-textual content. Similarly, the "Document loaders" module in LangChain exposes a "load" method for loading data as documents from a configured source.


\begin{center}
\begin{tcolorbox}[colback=gray!13,
                  colframe=black,
                  width=\columnwidth,
                  arc=1mm, auto outer arc,
                  boxrule=0.35pt,
                 ]
\textit{In summary, only LLM-SH can assist LLMs in handling unconventional input contents, while LLM-LNG and LLM-RSTR need to rely on LLM-SH to achieve this functionality. LLM-SH deals with content exceeding token limits by splitting, filtering, reassembling, or summarizing long documents. Regarding non-textual content, LLM-SH processes the text portions extracted from multimedia files or generates multi-modal multimedia files based on pure text content.}
\end{tcolorbox}
\end{center}

\vspace{2mm}
\textbf{Beneficial Effects of the Reasoning Process.}
The benefits of prompting frameworks for LLMs in the reasoning process are primarily evident in two capabilities: accelerating reasoning and ensuring that reasoning results meet preset requirements. In the current scenario with a surge in data volume and user usage, the speed of LLMs' reasoning is crucial for completing tasks efficiently. Additionally, improved reasoning results enable LLMs to accomplish more work in a given unit of time, greatly facilitating the application of these powerful natural language processing models in tasks requiring rapid responses and high throughput. For example, it's useful in online customer support (providing instant responses to customer queries for better user experiences), real-time financial analysis (for timely financial decision-making by analyzing news events and market data), and real-time sentiment analysis (for tracking product or brand feedback on social media in real-time).

\textit{Comparative Analysis.}
Regarding the crucial functionality of accelerating reasoning, most prompting frameworks do not provide support. Among the prompting frameworks surveyed, only LMQL within LLM-LNG offers this relevant service. LMQL accelerates inference by providing eager validation during LLM runtime. The principle behind eager validation is that LLMs typically generate text sequentially, similar to how humans write. When users make requests to LLMs with conditions like "the output must satisfy A and B," LMQL monitors the output as it's being generated. If LMQL detects that the currently outputted portion no longer satisfies condition A, it forcibly stops the LLM's execution and triggers it to re-infer, which eliminates the need to wait until the LLM completes the entire response to determine if the output complies with the conditions. For example, if a user instructs Chatgpt with the input "Write a poem about the sky, excluding clouds and birds," conditions A and B are "excluding clouds" and "excluding birds." If Chatgpt generates a description of "clouds" during the writing process, it no longer meets the user's request. Therefore, LMQL performs a forced termination of Chatgpt and resumes from the error-free portion to continue the task. This "validate-as-you-go" approach significantly accelerates LLM inference.

To ensure that inference results meet preset requirements, LLM-LNG, LLM-RSTR, and LLM-SH all utilize validate templates or allow for user-defined templates to ensure more standardized and accurate prompts (inputs). For instance, features like "Prompts" in LangChain's Model I/O module and "PromptTemplate" in Haystack enable predefined templates as well as user-defined templates within the framework. Additionally, the previously mentioned eager validation not only accelerates inference but also enforces stricter compliance with output requirements, making it easier to obtain outputs consistent with expectations and standards.


\begin{center}
\begin{tcolorbox}[colback=gray!13,
                  colframe=black,
                  width=\columnwidth,
                  arc=1mm, auto outer arc,
                  boxrule=0.35pt,
                 ]
\textit{In summary, to enhance the inference process of LLMs, LLM-LNG, LLM-RSTR, and LLM-SH all use methods such as predefined templates or compatibility with user-defined templates to ensure that inference results better meet requirements. Only LMQL within LLM-LNG supports the acceleration of the reasoning process during LLM runtime.}
\end{tcolorbox}
\end{center}

\vspace{2mm}
\textbf{Control of LLMs Outputs.}
Prompting framework's role in achieving controllable generation with LLMs primarily addresses two issues: imposing fine-grained structural constraints on generated output and ensuring that LLMs do not produce sensitive, non-compliant, or unsafe content.
The significance of structured output from LLMs lies in its ability to transform natural language text into a structured format, making information easier to process, analyze, and apply. The structured output aids in tasks such as extracting key information from text, building knowledge graphs, and performing data analysis. Structured data plays a crucial role in information management and data analysis, supporting decision-making, process optimization, insight discovery, and the development of intelligent applications. Structured data is essential in various industries and domains, helping organizations better understand and leverage their data assets.
The safety and relevance of LLMs' output content are crucial because these factors directly impact whether the text generated by the model is appropriate, accurate, and useful. For example, content filtering on social media platforms ensures that content posted on these platforms does not contain inappropriate or rule-violating material, maintaining the platform's reputation and user experience. In online customer support, automated response systems generate information only related to product support or common issues, ensuring that responses generated by automated systems are both on-topic and free from inappropriate content. Maintaining the online platform's brand reputation, protecting user rights, providing accurate information, and enhancing the availability of automated systems are all essential. Using prompting frameworks to assist in the automated and intelligent handling of LLMs' output content, ensuring that generated text meets specific requirements, has practical applications in social media, news media, e-commerce, online education, and many other fields. 

\textit{Comparative Analysis.} 
When it comes to constraining the output structure of LLMs, LLM-SH, LLM-LNG, and Content LLM-RSTR typically provide interfaces or pre-packaged functions that, when given custom structured templates provided by users, transform LLMs' output into the corresponding structured format. For example, LangChain's Model I/O module offers "Output parsers," which are classes designed to help structure language model responses. By invoking LangChain's built-in functions and combining them with user-configured structural templates, LLM-SH's LangChain directly supports various commonly used structured output forms, such as lists, time formats, enumerations, JSON, and more. In Semantic Kernel within LLM-SH, LMQL within LLM-LNG, and SudoLang, which are designed with frameworks and syntax that allow the mixing of natural language with code, support for structured data is achieved through user-configured custom structural templates. This means that users need to design, describe, and write their own templates in the corresponding framework's custom template locations. However, in the case of Structure LLM-RSTR, a category of prompting frameworks specifically designed for structured output, they offer built-in features for structured data generation and validation and excel at supporting formats like JSON and dialogue-based data. For instance, in Guidance, you can use the "{{gen}}" command to interleave generation, prompting, and logic control in a continuous sequential flow that aligns with how the language model processes text. In TypeChat, there's the provision of Schema, a data structure used to describe the expected format and fields of a prompt. By defining a Schema, you can validate and correct user input to ensure it adheres to the expected format and requirements.

Securing and ensuring compliance with the output content of LLMs is crucial. However, except for Content LLM-RSTR within LLM-RSTR, no other prompting frameworks address this issue. In Content LLM-RSTR, Guardrails is a Python package that adds type and quality assurance to LLMs' output, including semantic validation, such as checking for biases in generated text and errors in generated code and takes corrective measures when validation fails. Guardrails provides a file format (.rail) for executing "specifications" (user-specified structural and type information, validators, and corrective actions) on LLMs' output and offers a lightweight API call wrapper to implement these specifications on the output. NVIDIA's NeMo-Guardrails follows a similar design philosophy, ensuring control, safety, and security of large model language generation by limiting templates and themes in conversations. NeMo-Guardrails employs a custom language to implement a three-layer protection mechanism, including Topical GuardRail for topic-related questions, Safety GuardRail for ensuring controlled responses, and Security GuardRail to protect against malicious or abusive questions. However, this approach, while practical, also introduces potential challenges such as configuration redundancy, inflexibility, operational risks, and limitations on the original conversational capabilities of LLMs.


\begin{center}
\begin{tcolorbox}[colback=gray!13,
                  colframe=black,
                  width=\columnwidth,
                  arc=1mm, auto outer arc,
                  boxrule=0.35pt,
                 ]
\textit{In summary, LLM-SH, LLM-LNG, and LLM-RSTR all offer different ways to support control over the structure of LLMs' output, but only Content LLM-RSTR within LLM-RSTR provides constraints and validation for the safety and compliance of LLMs' output content, albeit with certain potential issues in its functionality.}
\end{tcolorbox}
\end{center}

\vspace{2mm}
\textbf{Impact of Invocation Costs.}
Reducing the cost of LLMs invocation primarily involves two approaches: firstly, reducing the frequency of invoking LLMs, and secondly, decreasing the number of tokens processed during LLMs invocation. Utilizing LLMs services on cloud computing platforms such as AWS, Azure, and Google Cloud incurs charges for each model invocation. These costs escalate with the frequency of invocations and the extent of computational resources used. One of the significant challenges in implementing LLMs for practical production scenarios is the high cost associated with invoking LLMs, which can be burdensome for individuals and even enterprises with limited financial resources. LLMs are typically billed based on the number of tokens generated or processed, with the number of tokens subject to charges determined by the prompt and the corresponding response length. The billing unit is the "token," which can represent a word, character, or punctuation symbol. In English, one token typically corresponds to one word, but for other languages, token forms may vary due to differences in language structures. Different models have varying token pricing standards; for example, GPT-3.5 is priced at \$0.002 per 1000 tokens, while GPT-4's token price is nearly six times higher than that of GPT-3.5. The maximum tokens supported and pricing of mainstream LLMs are shown in Tab. \ref{tab:cost}. While the cost per individual token may appear acceptable from a computational perspective, it becomes substantial for long-term usage and complex tasks. For instance, translating large documents, especially complex technical documents across multiple languages, may require extended processing time and additional computational resources. Similarly, employing LLMs for large-scale data analysis, text mining, or information extraction tasks, such as processing tens of thousands of news articles to extract key information, might necessitate distributed computing environments and substantial storage, resulting in high invocation costs.

\textit{Comparative Analysis.} 
In terms of reducing the frequency of LLMs invocation, LLM-SH introduces the "Memory" module that stores historical query information. Leveraging vector databases, enables a form of "learning from the past," so that when encountering the same or similar questions, LLMs do not need to be invoked again, thus saving unnecessary computational expenses. For instance, LLM-SH's LangChain implements a "Memory" module for storing past interactions, Haystack includes a "Module memory" module with the InMemoryDocumentStore class for recording and retrieving interaction content, and Griptape offers a "Conversation Memory" module with BufferConversationMemory functionality for constructing prompts with sliding window tasks. In contrast, LLM-LNG primarily reduces the frequency of invoking LLMs by embedding preprocessing programs and modules within the framework. Simple questions are first handled by smaller or free models, and then LLMs are utilized for more detailed processing, thus reducing the number of LLMs invocations. However, this method has limitations in broader tasks. Additionally, LLM-RSTR frameworks typically do not consider this functionality. Regarding the prompting framework's role in reducing the number of tokens that LLMs need to process, LLM-SH employs techniques like splitting, filtering, concatenating, or summarizing text segments to abbreviate and simplify the content to be processed. This aligns with the earlier-discussed approach of handling extremely long documents.


\begin{center}
\begin{tcolorbox}[colback=gray!13,
                  colframe=black,
                  width=\columnwidth,
                  arc=1mm, auto outer arc,
                  boxrule=0.35pt,
                 ]
\textit{In summary, the current state of prompting frameworks does not fully address the crucial challenge of reducing the cost of LLMs invocation. LLM-SH employs "Memory" functionality and text manipulation techniques to reduce the frequency of invocations and the number of tokens to be processed. LLM-LNG relies on embedded preprocessing modules, which have limited applicability, while LLM-RSTR frameworks generally do not implement this capability. Therefore, in the future, reducing the cost of LLMs invocation should be an essential area of development for prompting frameworks.}
\end{tcolorbox}
\end{center}

\vspace{2mm}
\textbf{Capability of Utilizing External Tools.}
The ability to use tools is one of the most significant distinctions between humans and other species. Currently, a crucial limitation of LLMs in practical applications is their inability to use external tools like humans. However, the advent of prompting frameworks effectively mitigates this deficiency. Typically, LLMs are generalized models, so the capability for LLMs to utilize external tools holds significant importance for enhancing their functionality, adapting to specific tasks and domain requirements, providing access to more data and resource support, and meeting compliance and security demands. For example, in custom text generation tasks, combining external plugins with LLMs allows the generation of industry-specific news reports, creative writing, or legal documents. In complex data query tasks, integrating LLMs with database query plugins enables them to respond to intricate business queries such as sales reports, inventory management, or medical record retrieval. Furthermore, multimodal processing plugins can be integrated with LLMs to analyze and generate content with text, images, and audio elements, such as social media posts or advertisements.

\textit{Comparative Analysis.} 
In terms of enabling LLMs to use external tools, LLM-SH outperforms LLM-RSTR and LLM-LNG due to its design focus. LLM-SH can support the integration of LLMs with a wide range of tools spanning various domains, such as web browsers, databases, email clients, image processing tools, speech recognition engines, code processors, and more. For instance, LangChain's Agents module offers a Toolkit function in which a collection of tools designed for specific tasks and conveniently loaded methods are available. These tools encompass Gmail, GitHub, SQL Databases, Vectorstore, and various Natural Language APIs. GripTape utilizes the Tools module's TextToolMemory to enhance the output of all tools, while also allowing users to perform different degrees of customization at the structural, task, or tool activity levels. TextToolMemory empowers LLMs to interact with the external world, generating non-textual content such as images, videos, PDFs, and others. In contrast, LLM-LNG and LLM-RSTR provide limited support for only the most commonly used external tools. For instance, LMQL in LLM-LNG offers services restricted to calculators and Wikipedia tools, while Guidance in LLM-RSTR offers integration with the Bing search engine.


\begin{center}
\begin{tcolorbox}[colback=gray!13,
                  colframe=black,
                  width=\columnwidth,
                  arc=1mm, auto outer arc,
                  boxrule=0.35pt,
                 ]
\textit{In summary, the current state of prompting frameworks for enabling LLMs to integrate with external tools is in its initial and somewhat immature stage. LLM-SH provides LLMs with the capability to use tools in relatively diverse scenarios to some degree, while LLM-LNG and LLM-RSTR offer very limited support in this regard.}
\end{tcolorbox}
\end{center}

\vspace{2mm}
\textbf{Maintenance of Historical Information.}
Maintenance of historical interaction information is crucial for LLMs as it provides them with the ability to store, retrieve, and reference information. This capability aids in better understanding context, maintaining coherence in complex tasks and long texts, and benefiting from past computations. However, apart from chat models like ChatGPT, which can record some history within the same conversation interface (with no sharing of information across different interfaces), LLMs typically process only the current context information provided in the prompt during service. This limitation is inconvenient for users as historical interaction information not only enhances interactions with LLMs, reducing the need for repetitive work but also stimulates non-dialogue LLMs like GPT-3.5 in chat tasks. Prompting frameworks address this need by offering "Memory" systems, supporting two fundamental operations: reading and writing. This enables the storage of historical interaction information and its utilization through queries, searches, and retrieval. Prompting frameworks empower LLMs with the ability to maintain historical interaction information, aiding in context understanding, information retrieval, multi-turn question-answering, document reading, error detection and correction using memory, caching intermediate computation results, and storing user preferences, historical behavior, or personalized information.

\textit{Comparative Analysis.} 
Regarding the maintenance of historical interaction information, LLM-SH excels due to its comprehensive Memory system, which allows integration with various tools across different domains. Taking LangChain as an example, the "Memory" can store past chat messages, queries, and results, effectively adding memory to the system. This storage function can be used independently or seamlessly integrated into various chains. Moreover, LangChain supports the use of embedded models and vector databases to store, query, and maintain historical information. During a single run, the system interacts with "Memory" twice, once when providing a query. The prompt includes two parts, one directly from user input, and the other possibly extracted from information stored in memory. The second interaction records the current interaction in memory for future queries and retrieval. LLM-SH's Memory system usually offers multiple ways to manage short-term memory, such as ConversionMemory and VectorStore-backed Memory. ConversionMemory summarizes ongoing conversations and stores the current summary in memory, passing it as short-term memory to LLMs in new rounds of conversation. This compression of historical conversation information is particularly useful for lengthy conversations. On the other hand, VectorStore-backed Memory stores all past conversations in vector form in a vector database. In each new conversation round, it matches the user's input against the vector database to find the most similar K sets of conversations. LLM-LNG and LLM-RSTR typically achieve historical information maintenance by invoking the corresponding functional modules in LLM-SH. An interesting feature in LLM-LNG is LMQL's "Caching Layer," implemented as a tree-like data structure. This layer caches all model outputs, including logs and historical information, to enable more efficient exploration of the LLM's token space during runtime, even in the presence of multiple variables, constraints, and tool enhancements. The cache can be seen as a tree with only appendices, explored during query execution and expanded based on query code, constraint conditions, and inferred execution scenarios.


\begin{center}
\begin{tcolorbox}[colback=gray!13,
                  colframe=black,
                  width=\columnwidth,
                  arc=1mm, auto outer arc,
                  boxrule=0.35pt,
                 ]
\textit{In summary, LLM-SH provides a relatively complete and comprehensive memory system for maintaining historical interaction information, while LLM-LNG and LLM-RSTR typically achieve this by invoking corresponding functional modules within LLM-SH.
}
\end{tcolorbox}
\end{center}

\subsection{\textbf{Documentation and Support.}}
Document quality and community support are crucial for a prompting framework. Document integrity refers to the accuracy, readability, and completeness of technical documentation, which assesses whether the development team provides detailed, clear, and comprehensive documentation to facilitate user learning and usage. Community support involves the presence of an active developer community around the framework, allowing users to receive timely assistance and feedback.

\vspace{1mm}
\textbf{Document Completeness.} 
The investigated prompting frameworks offer three primary deployment methods. The first method is command-line installation, where Python-based frameworks use commands like "pip install" or "conda install," and JavaScript-based frameworks use "npm install" for deployment. This method is available for almost all prompting frameworks. The second method is a custom interactive user interface, providing a playground that allows users to interact with the framework directly through a user-friendly interface without the need for installation, for example, LMQL in LLM-LNG. The third method is the "open-and-use" approach, where deployment and execution are done directly on the interface provided by LLMs, offering the most straightforward and simple operation. SudoLang in LLM-LNG is an example. Overall, the prompting frameworks studied provide comprehensive technical documentation that covers almost all features and use cases. They also offer detailed example code, tutorials, and operation guides, making it easy for users to get started, understand, and use these frameworks. These technical documents are readily accessible on GitHub. Additionally, some frameworks have created well-designed web pages with high readability and attractiveness, such as LangChain in LLM-SH.

\vspace{1mm}
\textbf{Document Readability.}
At present, the technical documentation of existing prompting frameworks is comprehensive and detailed but may lack readability. The rapid proliferation of LLMs and their derivatives has led to a need for frequent updates to the technical documentation. These updates may occur daily to accommodate more products and features. The introduction of new features can sometimes render old ones unusable or introduce bugs, causing significant challenges for users and developers. Furthermore, the continuous addition of features and product introductions can make the documentation structure complex and the content increasingly lengthy. This can lead to a situation where the documentation becomes overwhelming and resembles a "code mountain" rather than maintaining the clarity and simplicity of its initial state, which can be frustrating for users. In summary, the dynamic nature of LLMs and their associated frameworks necessitates frequent documentation updates, but these updates must be managed carefully to maintain clarity and usability for users and developers.

\vspace{1mm}
\textbf{Community Support.}
Regarding community support and user feedback channels, the development teams behind these prompting frameworks typically offer Discord online community services. Users can ask questions, share experiences, and engage with other users in these communities. Furthermore, they maintain official support teams on Twitter and have dedicated communication topics (tags) for users to provide feedback, suggestions, and opinions, thus facilitating framework improvements. Additionally, they provide communication spaces on GitHub for users and developers.

\subsection{Challenges Confronting the Prompting Framework} 
The existing prompting frameworks currently alleviate many limitations of LLMs in practical applications, significantly enhancing the capabilities of LLMs themselves and further elevating their performance. However, burgeoning development still encounters numerous challenges. In this section, we will analyze the challenges and opportunities that the prompting framework currently faces in terms of functionality implementation and issues related to safety and ethics.

\subsubsection{\textbf{Security Mechanisms of Prompting Framework.}}
For a developed tool, both functionality and security mechanisms are equally crucial. Software that cannot guarantee the security of user information and is not harmless to society, regardless of its robust functionality, is unlikely to gain user support. Existing evidence suggests that LLMs and relatively mature prompting frameworks like LangChain have certain security issues \cite{pedro2023prompt, barrett2023identifying, liu2023prompt, mei2023notable, wei2023jailbroken}. However, the existing prompting frameworks have paid minimal attention to the significance of security ethics and privacy protection. Due to LLMs possessing an "Achilles' heel" – being uncontrollable generative AI, we cannot predict the outputs, which can be fatal in certain scenarios. Therefore, we argue that the security mechanisms that prompting frameworks urgently need to enhance should consist of two parts: defense against prompt-based attacks and safeguarding the behavior of LLMs.

\vspace{1mm}
\textbf{Defense Against Prompt-based Attacks.}
We begin by introducing the concept of Prompt-based Attacks and then provide some suggestions to mitigate this issue within future prompting frameworks. Prompt-based Attacks share essential similarities with Prompt Engineering in that they both aim to obtain desired outputs through expertly crafted, rational, and optimized instructions. However, Prompt Engineering is user-oriented, while Prompt-based Attacks adopt a hacker-attack perspective. Malicious inputs from external sources can contaminate the model's outputs through Prompt-based Attacks, thereby exerting influence on external systems, resulting in adversarial actions. The impact of such attacks depends on the capabilities granted to the model by the system. Prompt-based Attacks refer to attempts to generate content that contradicts the developer's intent using LLMs \cite{ignore_previous_prompt, greshake2023more, liu2023prompt, deng2023jailbreaker, yan2023virtual}. Typically, there are two forms of attacks: Prompt-based Deception (bypassing scrutiny through linguistic techniques) and Prompt-based Injection (tampering with instructions). In scenarios limited to content generation, the harm caused by these attacks may be relatively insignificant. However, with the proliferation of various prompting frameworks and projects like AutoGPT, more individuals are granted execution authority over LLMs, expanding the potential danger. These scenarios encompass not only the creation of email worms utilizing automated email processing functions but also poisoning email extraction systems through web-based attacks, code poisoning through code completion mechanisms, and more ominous possibilities, such as manipulating or accessing local files or diverting funds if a private prompting framework assistant were to be compromised or granted execution authority.
    
As for defense mechanisms against Prompt-based Attacks, we offer several suggestions from the perspective of instruction design. Firstly, when constructing prompts for interaction with LLMs, it is advisable to employ delimiters to rigorously differentiate instructions from content, which involves encapsulating user-generated content within delimiters and imposing specific authority constraints, a practice endorsed by OpenAI and machine learning expert Andrew Ng as a best practice. Secondly, place crucial instructions at the end of the prompt. Given that most malicious instructions currently tend to disregard preceding instructions, positioning developer instructions at the end of the entire prompt is indeed a straightforward and convenient method in practice. Lastly, consider incorporating a pre-filtering layer, such as establishing whitelists and blacklists for prompt content. Whitelists define permissible inputs, for instance, in the case of designing a machine translation model from Chinese to English, pre-detect and allow only Chinese characters in the prompt for synthesis with the model. Conversely, blacklists prohibit specific inputs, for instance, detecting common jailbreak-related phrases or instructions like "ignore" and refrain from inputting them into the model or checking for the presence of phrases prohibited by other legal regulations.

\vspace{1mm}
\textbf{Safeguards of LLMs' Behavior.}
As is widely acknowledged, LLMs while gaining favor from millions of users, occasionally exhibit undesirable behaviors such as escaping, hallucinating, and deception. Therefore, it is crucial to "muzzle" LLMs by adding an additional safeguard layer to their output. Starting from the design principles and implementations of the prompting framework, augmenting LLMs with an additional protective mechanism within the prompting framework is both reasonable and aligned with software development principles. This augmentation allows for the early interception and modification of objectionable or non-compliant content in the output of LLMs before obtaining the final results.

The Nemo Guardrails developed by Microsoft, as mentioned earlier, still has several issues when it comes to protecting LLMs' behavior. Firstly, it uses its proprietary definition language, which may appear more user-friendly but limits its extensibility. Moreover, it transforms a model's security and control problem into a manual policy configuration problem. This approach brings various potential issues such as redundant configurations, lack of flexibility, operational risks, and potential curtailment of the original capabilities of large language models. It seems to be more like a compromise solution in today's commercial scenarios where there is a desire to effectively utilize the generation capabilities of large models but no effective solution for controllability and security.
    
Therefore, we believe that future prompting frameworks should be designed based on mainstream programming languages, employing advanced technologies to establish flexible and automated mechanisms for extracting and configuring protection rules, which focus on three main aspects: topic relevance, content safety, and application security, ultimately standardizing the behavior of LLMs. Thematic integrity safeguards aim to prevent LLMs from going off-topic. LLMs possess a richer imagination and are more capable of creative code and text generation compared to other AIs. However, for specific applications such as coding or customer service, users do not want them to "stray from the intended scope" while addressing issues, and generating irrelevant content. In such cases, topic-constrained safeguards are required. When a large model generates text or code that goes beyond the predefined topic, these safeguards guide it back to the designated functionality and topic. Content safety safeguards are intended to prevent incoherent output from large models. Incoherent output includes two scenarios: factual inaccuracies in the answers generated by LLMs, which are things that "sound reasonable but are entirely incorrect," and the generation of biased or malicious output, such as using offensive language when prompted by users or generating unethical content. Application security refers to restricting the application's connections to known secure third-party applications. The prompting framework should avoid exposing LLMs to malicious attacks from external sources during task execution. This includes preventing the induction of LLMs to call external virus plugins and defending against hackers who may attempt to attack LLMs through methods like network intrusion or malicious software.

\subsubsection{\textbf{Capability Limitations of Prompting Framework}}
In comparing and analyzing the capabilities and features of prompting frameworks, we delineate 6 dimensions. Prompting frameworks exhibit commendable performance across these capability dimensions, but they still exhibit shortcomings in the degree of implementation within each capability dimension. For instance, concerning the handling of unconventional inputs, current prompting frameworks can significantly assist LLMs in text processing leaps but remain somewhat constrained in multi-modal scenarios (e.g., video and images). Similarly, limitations exist in invoking external expert models. Furthermore, as the landscape of large models continuously evolves with the emergence of numerous LLMs and external plugins, many prompting frameworks adapt by continually adding encapsulated invocation functions to support relevant services. However, this rapid evolution introduces several challenges. The updated functionalities are often haphazardly aggregated and do not offer the same level of support as native capabilities provided by the framework. Additionally, more intricate functionalities may lead to contradictions or duplications when not properly planned, resulting in a steep learning curve for users. Consequently, we analyze the limitations of current prompting frameworks from these perspectives: an increasingly steep learning curve, constraints in invoking external interfaces and handling multi-modal I/O.

\vspace{1mm}
\textbf{Increasingly Steep Learning Curve.}
With the explosive growth in the field of large models, numerous related tools and plugins have emerged, which has posed significant challenges to the development and maintenance of prompting frameworks, requiring software engineers to quickly familiarize themselves with these emerging external tools and integrate their interfaces into the existing prompting framework. This has resulted in some functionality stacking and redundancy. For example, native functionalities within the prompting framework typically allow for straightforward invocation with uniform interfaces, achieved by simply modifying parameters. Conversely, newly added functionalities often necessitate users to invoke other packages and employ relatively complex and non-uniform calling procedures. Consequently, users must acquire additional knowledge before usage, resulting in steeper learning curves. The relatively complex processes and methods also lead to suboptimal user experiences or program bugs. Furthermore, the introduction of new functionalities may disrupt the proper functioning of previously implemented native features, which is a recurring issue within existing prompting frameworks. As previously indicated, it has been observed that the technical documentation of the existing prompting framework exhibits deficiencies in terms of readability. The emergence of this problem is logical, and an effective mitigation strategy involves developers considering future developments and changes in the framework's structural design, which entails enhancing the modularity, scalability, and standardization of the prompting framework.

\vspace{1mm}
\textbf{Constraints in Invoking External Interfaces.}
Currently, prompting frameworks can support relatively basic external tool usage, such as browsing the web or querying databases. However, we argue that the accessibility of external expert models remains rudimentary within most prompting frameworks. For instance, current prompting frameworks cannot fully assist LLMs in utilizing the widely-used "Microsoft Suite" in commercial and office environments, including Microsoft Word, Microsoft Excel, and Microsoft PowerPoint. Additionally, prompting frameworks lack secure protocols for accessing and handling users' private or commercial data that meet security and privacy requirements. Furthermore, because current prompting frameworks can only handle text-based tasks, there are limitations in handling multi-modal inputs such as videos, Word documents, emails, etc. This makes it challenging to provide support for widely-used functional tools like YouTube (one of the world's largest video-sharing platforms with billions of users), arXiv (a significant open-access academic preprint platform), Twitter (a prominent social media platform in the realm of social media and news dissemination), Outlook (one of Microsoft's widely-used email and calendar management tools), and others. In the future, prompting frameworks should aim to not only assist LLMs in supporting these widely-used mainstream software but also allow for the integration of more emerging or niche tools or platforms to foster a more vibrant AI community ecosystem.

\section{future directions and conclusion}
\label{sec:con}

\subsection{Conclusion of Existing Prompting Frameworks}
In this paper, we elucidate the genesis of the prompting framework and its underpinning technological foundations. Subsequently, we proffer a conceptual definition of the prompting framework, along with its requisite characteristics of modularity, abstraction, extensibility, and standardization. We then classify the prompting framework based on usage scenarios and technical attributes into three categories: the shell of  LLMs (LLM-SH), language for interaction with LLMs (LLM-LNG), output restrictors of LLMs (LLM-RSTR). Following this categorization, we conduct a comprehensive comparative analysis of the compatibility, capabilities and features, documentation, and community support of these prompting frameworks across various dimensions. Finally, we delineate the challenges currently confronting the development of prompting frameworks. Additionally, we introduce several practical relevant prompt-based tools that fall outside the purview of the prompting framework domain and tools that play a significant role in assisting LLMs in accomplishing tasks. In this section, we summarize the applicability and limitations of existing prompting frameworks.

Despite the various attempts made by current prompting frameworks to alleviate the limitations of LLMs in real-world applications, there are still challenges and limitations that need to be addressed. These frameworks have emerged with different focuses and features, addressing various dimensions of user concerns, including documentation and community support, compatibility, capabilities (such as the ability to use external tools, cost reduction, etc.), which have indeed made strides in solving some of the issues. However, it is important to note that current prompting frameworks can be considered as compromise solutions to meet user needs in today's commercial scenarios, rather than fully future-proof methods. The limitations of existing prompting frameworks primarily revolve around the lack of support for security mechanisms and inherent limitations in their capabilities.

\vspace{1mm}
\textbf{Security Mechanisms.} Regarding the security mechanisms within prompting frameworks, including resistance to prompt-related attacks and constraints on LLMs' behavior, there are currently significant limitations. Firstly, most prompting frameworks do not adequately address resistance against prompt-related attacks, such as injection and deception. These vulnerabilities pose a severe threat to system security. Additionally, in terms of constraining LLMs' behavior, current prompting frameworks rely primarily on manually configured safety policies similar to Reinforcement Learning from Human Feedback (RLHF). They have not fully leveraged advanced technologies and methods, which can result in issues like redundant configurations, inflexibility, operational risks, and even limitations on the original functionality of LLMs.

\vspace{1mm}
\textbf{Capability Limitations.} The limitations in the capabilities of prompting frameworks themselves are evident, especially when it comes to developing applications with large language models (LLMs). One of the primary reasons for this limitation is that many of the issues in LLM applications are rooted in the deficiencies of the underlying technology of large models, emphasizing the importance of prompt engineering. For instance, when manipulating LLMs to perform highly complex tasks using prompting frameworks, developers often rely on highly customized, handcrafted prompts. However, many existing prompting frameworks are designed with a "simplification to complexity" principle in mind. They assume that more complex structures lead to more comprehensive functionality. In other words, these frameworks tend to be overly complex and do not provide sufficient openness in terms of prompt design. As a result, users often find themselves needing to configure many aspects of the system themselves, but prompting frameworks do not provide appropriate support for this requirement. Therefore, simplifying and streamlining the configuration process and providing more open and flexible prompt design options could enhance the usability and effectiveness of these frameworks in developing complex LLM applications. Simultaneously, the current prompting framework still faces notable shortcomings in its accessibility to the broader external world, such as the limited support for third-party tools, including multimodal tools, and mainstream platforms like arXiv and Twitter.

\subsection{Future Directions}
We believe that the next-generation of prompting frameworks should overcome the limitations mentioned above while integrating the strengths of the three prompting frameworks in this paper, which can provide users with more concise and compact interaction channels, facilitate LLMs interactions with powerful third-party interfaces, and enable interactions with higher-quality and more tailored. The next-generation prompting framework should be a comprehensive platform that is more streamlined, more secure, more versatile, and more standardized, which seamlessly integrates development, testing, evaluation, maintenance, expansion, and communication with LLMs, constituting an organic LLM ecosystem.

\vspace{1mm}
\textbf{More streamlined.} The next-generation prompting framework should embody a higher degree of streamlining, primarily manifested in the simplification of user interactions with LLMs and the interactions between LLMs and environments. Furthermore, the technical architecture and documentation should exhibit enhanced compatibility with new products and technologies, coupled with a more user-friendly learning curve and instructional materials. In essence, it should adhere to the principle of "simplify without oversimplifying, embracing the concept of simplicity as the ultimate sophistication."

\vspace{1mm}
\textbf{More secure.} The next-generation prompting framework ensures the secure and compliant generation of content by Large Language Models (LLMs) while safeguarding user privacy and security, serving as a bidirectional security barrier between users and LLMs and between LLMs and applications.

\vspace{1mm}
\textbf{More versatile.} The next-generation prompting framework seamlessly integrates with more diverse, feature-rich external applications, enabling LLMs to excel in various domains such as healthcare, research, education, transportation, and more.

\vspace{1mm}
\textbf{More standardized} The next-generation prompting framework adheres to established domain standards, which are widely accepted sets of rules, guidelines, specifications, or best practices within specific domains to ensure the quality, consistency, and reliability of products, services, or processes. Examples include ISO 27001 for information security management systems (ISMS) in the field of information technology and GMP standards for quality management in pharmaceutical and medical device manufacturing in the healthcare domain. These standards facilitate compliance across different prompting frameworks, eliminating the need for users to acquire additional knowledge and promoting mutual support and complementarity between different prompting frameworks.

\vspace{1mm}
\textbf{Organic LLMs ecosystem.} The next-generation prompting framework seamlessly integrates with LLMs, serving as a comprehensive platform for LLM development, testing, comparison, evaluation, user interaction, and developer communication. This integration fosters an ecosystem that evolves through continuous feedback, ultimately delivering enhanced services and enabling leaps in technology and application development.

\bibliographystyle{ACM-Reference-Format}
\bibliography{ref}

\end{document}